\documentclass{aa}  
\usepackage{graphicx} 
\usepackage{natbib}
\bibpunct{(}{)}{;}{a}{}{,} 
\usepackage{txfonts}
\usepackage{longtable,lscape}
\usepackage{color}
\begin{document}
\title{Multi-frequency measurements of the NVSS foreground sources in
  the Cosmic Background Imager fields}

\subtitle{I. Data release}

\author{E.~Angelakis\inst{1},
  A.~Kraus\inst{1}, 
  A.~C.~S.~Readhead\inst{2}, 
  J.~A.~Zensus\inst{1}, 
  R.~Bustos\inst{3,4}, 
  T.~P.~Krichbaum\inst{1}, 
  A. Witzel\inst{1} 
  \and
  T.~J.~Pearson\inst{2} 
}

\offprints{E. Angelakis}

\institute{Max-Planck-Institut f\"ur Radioastronomie, 
  Auf dem H\"ugel 69, 53121 Bonn, Germany
  \and
  California Institute of Technology, 1200 East California Boulevard, California 91125, Pasadena, USA
  \and
  Universidad de Concepci\'{o}n, Casilla 160-C, Concepci\'{o}n, Chile 
  \and
  University of Miami, Department of Physics, 1320 Campo Sano Drive, FL 33146, USA\\
}

\date{Received -, -; accepted -, -}


\abstract 
{We present the results of the flux density measurements at 4.85\,GHz
  and 10.45\,GHz of a sample of 5\,998 NVSS radio sources with the
  Effelsberg 100\,m telescope .}
{The initial motivation was the need to identify the NVSS radio
  sources that could potentially contribute significant contaminating
  flux in the frequency range at which the Cosmic Background Imager
  experiment operated.}
{An efficient way to achieve this challenging goal has been to compute
  the high frequency flux density of those sources by extrapolating
  their radio spectrum. This is determined by the three-point spectral
  index measured on the basis of the NVSS entry at 1.4\,GHz and the
  measurements at 4.85\,GHz and 10.45\,GHz carried out with the 100\,m
  Effelsberg telescope.}
{These measurements are important since the targeted sample probes the
  weak part of the flux density distribution, hence the decision to
  make the data available.}  
{We present the table with flux density measurements of 3\,434 sources
  that showed no confusion allowing reliable measurements, their
  detection rates, their spectral index distribution and an
  interpretation which explains satisfactorily the observed
  uncertainties.}  

\authorrunning{Angelakis et al.}
\titlerunning{Multi-frequency measurements of the NVSS foreground sources in
  the CBI fields}

\maketitle

\section{Introduction}
\label{sec:intro}
Targeted multi-frequency surveys can be very efficient in serving
several fields of astrophysical research such as revealing new GPS and
HFP sources, estimating higher frequency source counts from
extrapolating the radio spectra and hence computing the confusion
limits etc. Consequently, they can be of essential importance in the
study of the Cosmic Microwave Background radiation (CMB) through the
characterization of the foregrounds. Here, we present the results of
the study of a sample of 5\,998 NRAO VLA Sky Survey
\citep[NVSS,][]{Condon1998AJ} sources at three frequencies. 1.4\,GHz
is provided by the NVSS catalog and 4.85\,GHz and 10.45\,GHz were
observed with the Effelsberg 100\,m radio telescope. The measurements
were initially motivated by the need to estimate the emission that
they could contribute at 31\,GHz. This is the band in which the Cosmic
Background Imager \citep[CBI,][]{Padin2001ApJ} operates, as explained
below. In a future publication we plan to use the extrapolated flux
densities in order to compute the source counts and the confusion
limits at higher frequencies and compare the results with those from
other surveys.

\subsection{The CMB contaminants}
\label{subsec:contaminants}
Having traveled the path between the surface of last scattering and
the observer, the CMB is subject to the influence of numerous sources
of secondary brightness temperature fluctuations, cumulatively
referred to as {\sl foregrounds}. The reliability of the information
extracted from the study of the primordial fluctuation power spectrum
is tightly bound to how carefully such factors have been accounted
for.

The potential contaminants can crudely be classified in those of
galactic and those of extragalactic origin \citep[for a review,
see][]{Refregier1999ASPC,Tegmark2000ApJ}. Moreover, depending on their
character, they influence the power spectrum at different angular
scales.  Galactic foregrounds could be the {\sl diffuse synchrotron
  emission} \citep[for a review, see][]{Smoot1999ASPC} attributed to
galactic relativistic electrons, the {\sl free-free emission}
originating in \ion{H}{ii} regions and the {\sl dust emission} due to
dust grains in the interstellar medium that radiate in a black body
manner. Extragalactic foregrounds could be the {\sl thermal} and the
{\sl kinematic Sunyaev-Zel'dovich effect} in galaxy clusters
\citep{Sunyaev1970ApnSS}, manifested through the distortion of the
black body CMB spectrum induced by either hot ionized gas in the
cluster (in the former case), or matter fluctuations (in the latter
case).

In the latter class, radio galaxies and quasars cumulatively referred
to as {\sl point radio sources}, populating the entire radio sky,
comprise by definition the most severe contaminant affecting small
angular scales. The sample studied in the current work is exactly the
NVSS point radio sources that lie within the fields targeted by the
CBI experiment.

\subsection{The Cosmic Background Imager}
\label{subsec:CBI}
The CBI is a 13-element planar synthesis array operating in 10 1-GHz
channels between 26 and 36\,GHz \citep{Padin2001ApJ}. It is located at
an altitude of roughly 5\,080\,m near Cerro Chajnantor in the Atacama
desert (northern Chilean Andes).  Its task was to study the primordial
anisotropies at angular scales from ~5\arcmin to 0.5\degr \citep[$400<
\ell<3\,500$, ][]{Padin2002PASP}. The observations of the primordial
anisotropies are made in four distinct parts of the sky, separated
from one another by 6\,hr in RA
\citep[][table~\ref{tab:fields}]{Mason2003ApJ,Pearson2003ApJ}.

From the NVSS catalog, it is known that within the CBI fields, there
are in total 5\,998 discrete radio sources with
$S_{1.4}\ge3.4$\,mJy. Inevitably, they comprise the potential
contaminants that may impose secondary fluctuations in the observed
background temperature field \citep{Readhead2004ApJ}.

\subsection{The Solution} 
\label{subsec:the_solution}
Instead of removing all potentially contaminated pixels from the CMB
maps (which would unavoidably cause a significant data loss), it would
suffice to identify the sources that contribute negligible flux
density at higher frequencies (below the few-mJy threshold) and ignore
them during the CMB data analysis. On the basis of the assumptions
that:
\begin{enumerate}
\item the radio spectrum is described by a simple power law of the
  form $S\sim \nu^{\alpha}$ (with $\alpha$ hereafter being the
  spectral index)
\item the spectrum is not time variable
\end{enumerate}
this identification can in principle be done by the extrapolation of
the radio spectrum as obtained at lower frequencies. The radio
spectrum consists of the flux density at 1.4\,GHz as extracted from
the NVSS and those at 4.85 and 10.45\,GHz as measured with the
Effelsberg telescope.

\subsection{The Sample} 
\label{subsec:the_sample}
\begin{figure}
  \centering
  \includegraphics[width=0.4\textwidth,angle=-90]{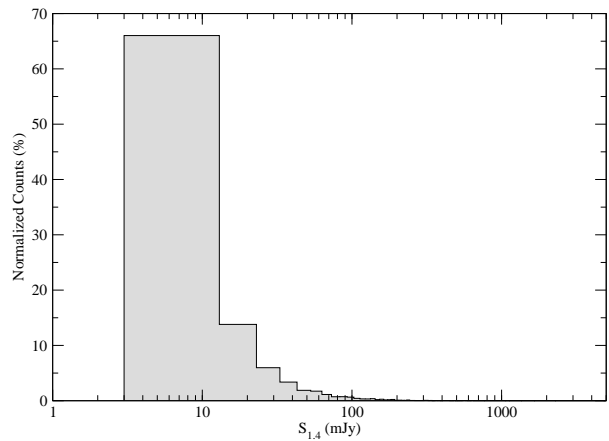}
  \caption{The NVSS 1.4\,GHz flux density distribution of our
    sample. Roughly 80\% is below 20\,mJy. This plot demonstrates
    clearly the choice made of radio ``quiet'' sky regions.}
  \label{fig:21distrib}
\end{figure}

The list of targeted sources includes all 5\,998 NVSS sources present
in the CBI target fields displaying $S_{1.4}\ge 3.4$\,mJy. At this
limit the NVSS is characterized by 99\% completeness. The sources are
distributed in four sky regions -- rectangular in ra-dec space -- with
their coordinates shown in table \ref{tab:fields}.
\begin{table}[t]
  \caption{
    The coordinates of the points defining the targeted fields.}
  \label{tab:fields}  
  \centering                    
  \begin{tabular}{c r@{}l r@{}l r@{}l r@{}l c} 
    \hline\hline                 
    Field &\multicolumn{2}{c}{NE} &\multicolumn{2}{c}{NW} &\multicolumn{2}{c}{SW} &\multicolumn{2}{c}{SE} &Area        \\    
          &\multicolumn{8}{c}{} &(deg$^{2}$) \\    
    \hline                        
    02$^{\mathrm{h}}$ &03$^\mathrm{h}$&$+$2.0$^\mathrm{d}$    &02.65$^\mathrm{h}$&$+$2.0$^\mathrm{d}$    &02.65$^\mathrm{h}$&$-$5.5$^\mathrm{d}$  &03$^\mathrm{h}$&$-$5.5$^\mathrm{d}$ &$41$\\
    08$^{\mathrm{h}}$ &09$^\mathrm{h}$&$+$0.0$^\mathrm{d}$    &08.60$^\mathrm{h}$&$+$0.0$^\mathrm{d}$    &08.60$^\mathrm{h}$&$-$5.5$^\mathrm{d}$  &09$^\mathrm{h}$&$-$5.5$^\mathrm{d}$ &$33$ \\
    14$^{\mathrm{h}}$ &15$^\mathrm{h}$&$+$0.5$^\mathrm{d}$    &14.60$^\mathrm{h}$&$+$0.5$^\mathrm{d}$    &14.60$^\mathrm{h}$&$-$7.0$^\mathrm{d}$  &15$^\mathrm{h}$&$-$7.0$^\mathrm{d}$ &$45$ \\
    20$^{\mathrm{h}}$ &21$^\mathrm{h}$&$-$2.0$^\mathrm{d}$    &20.60$^\mathrm{h}$&$-$2.0$^\mathrm{d}$    &20.60$^\mathrm{h}$&$-$8.0$^\mathrm{d}$  &21$^\mathrm{h}$&$-$8.0$^\mathrm{d}$ &$36$ \\ 
    \cline{10-10}
                     &\multicolumn{2}{c}{} &\multicolumn{2}{c}{} &\multicolumn{2}{c}{} &\multicolumn{2}{c}{} & 155 \\    
    \hline                                  
  \end{tabular}
  \begin{list}{}{}
    \item Each point is given by its J2000 (ra, dec) with right ascension
    (ra) measured in hours and the declination (dec) in degrees.
  \end{list}
\end{table}
The criterion for the selection of these regions has been their
foreground emission. It has been required that they have {\sl IRAS}
100\,$\mu$m emission less than 1 MJy sr$^{-1}$ \citep{Pearson2003ApJ},
low galactic synchrotron emission and no point sources brighter than a
few hundred mJy at 1.4\,GHz. It is clear therefore that the sample of
the 5\,998 sources represents the weak part of the flux density
distribution. This is the most prominent characteristic of the
sample. In fact, it is readily shown in Fig.~\ref{fig:21distrib} that
roughly 80\% of those sources are of $S_{1.4}\le20$\,mJy.
The large galactic latitudes of the targeted fields indicate that the
sample is likely to consist completely of extragalactic discrete radio
sources that is, quasars and radio galaxies.

In the current work we present the results extracted from a sample of
3\,434 sources which as it is discussed in
Sect.~\ref{subsec:confflavors} and \ref{sect:measuredfluxdensities}
show no confusion from field sources and hence allow reliable
measurements.

\section{Observations}
\label{sec:ObsDataRed}

\subsection{Observing System}
\label{subsec:System}
The flux density measurements were conducted with the Effelsberg
telescope between July 2003 and July 2006. The multi-beam heterodyne
receivers at 4.85\,GHz and 10.45\,GHz were used. Multi-feed systems
use ``software beam-switch'' for removing mostly linear troposphere
effects. Each receiver was used in two-beam mode (although the
10.45\,GHz one is equipped with 4 feeds). In both cases, the beams are
separated in azimuth and each delivers left-handed and right-handed
circular polarisation channels (LCP and RCP respectively).  Both
systems are mounted in the secondary focus cabin of the 100\,m
telescope. Table~\ref{tab:receivers} gives their characteristics.
\begin{table}[b]
  \caption{
    Receiver characteristics.}     
  \label{tab:receivers}  
  \centering                    
  \begin{tabular}{ccccccc} 
    \hline\hline                 
    $\nu$ &$\Delta\nu$ &$T_\mathrm{sys}$ &$\Gamma$  &$\theta$  &$X$    &Pol. \\    
    (GHz) &(MHz)     &(K)              &(K/Jy)       &(\arcsec) &(\arcsec) &     \\    
    \hline                        
     4.85 &500 &27  &1.5 &146  &485   &LCP, RCP \\     
    10.45 &300 &50  &1.3 &67   &182.4 &LCP, RCP \\
    \hline                                  
  \end{tabular}
  \begin{list}{}{}
  \item Column 1, is the central frequency while Column 2 is the
    receiver bandwidth. Column 3, gives the system temperature and
    Column 4 the sensitivity. Column 5, shows the full width at half
    maximum and Column 6 the angular separation between the two
    beams. Finally, in Column 7 we give the polarization channels
    available.
  \end{list}
\end{table}

\subsection{Observing Technique}
\label{subsec:Technique}
In order to achieve time efficiency, the observations have been made
with the ``on-on'' method.  Its essence relies on having the source in
either of the two beams at each observing phase whereas the other feed
is observing the atmosphere off-source (the angular distance of the
used feeds is given in table~\ref{tab:receivers}). The subtraction of
the two signals removes linear atmospheric effects. For clarity, one
complete measurement cycle will hereafter be termed as one {\sl
  scan}. Each scan, in our case, consists of four stages, or {\sl
  sub-scans}.

\begin{figure}
  \centering
  \includegraphics[width=0.4\textwidth]{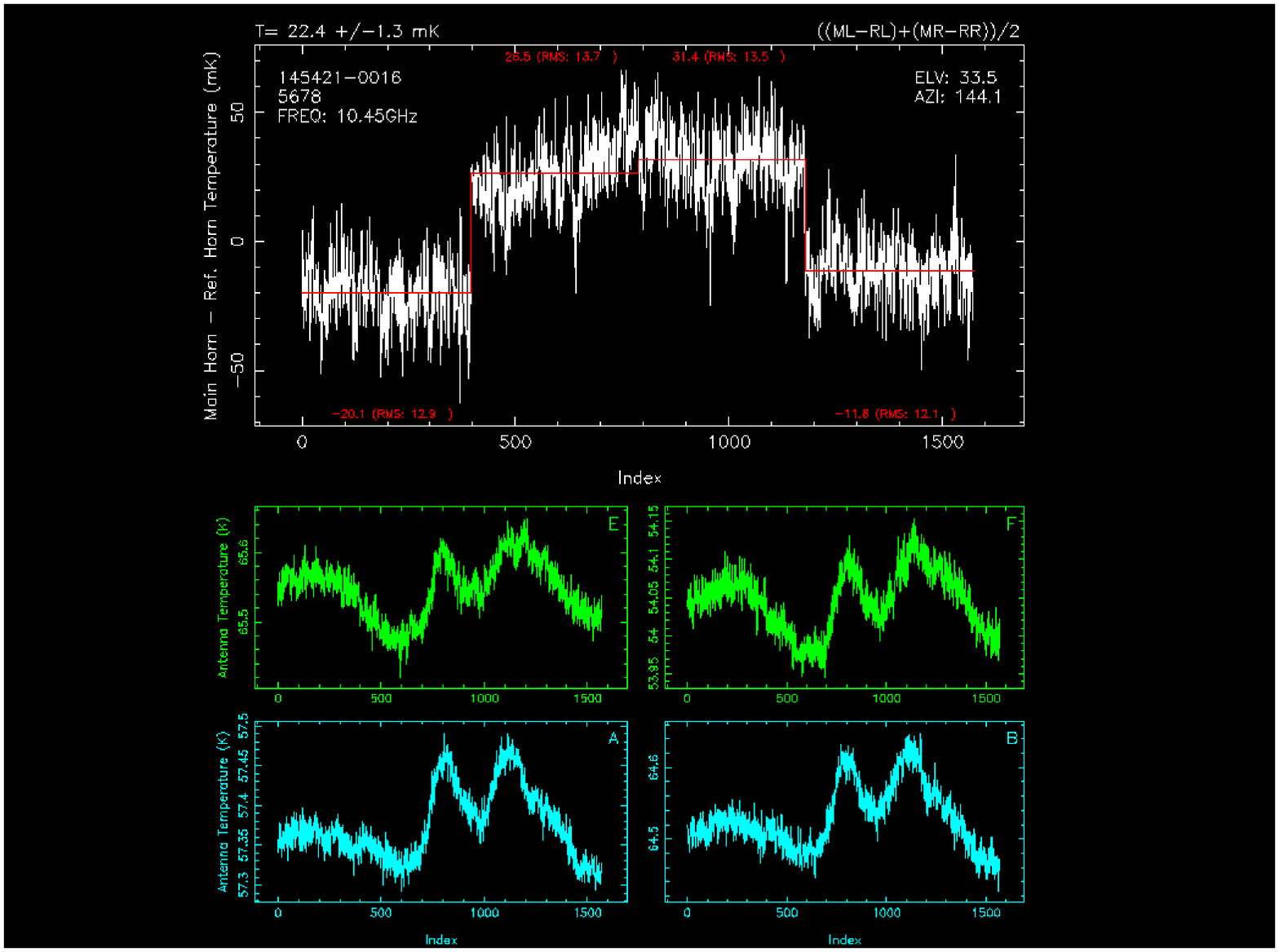}
  \includegraphics[width=0.4\textwidth]{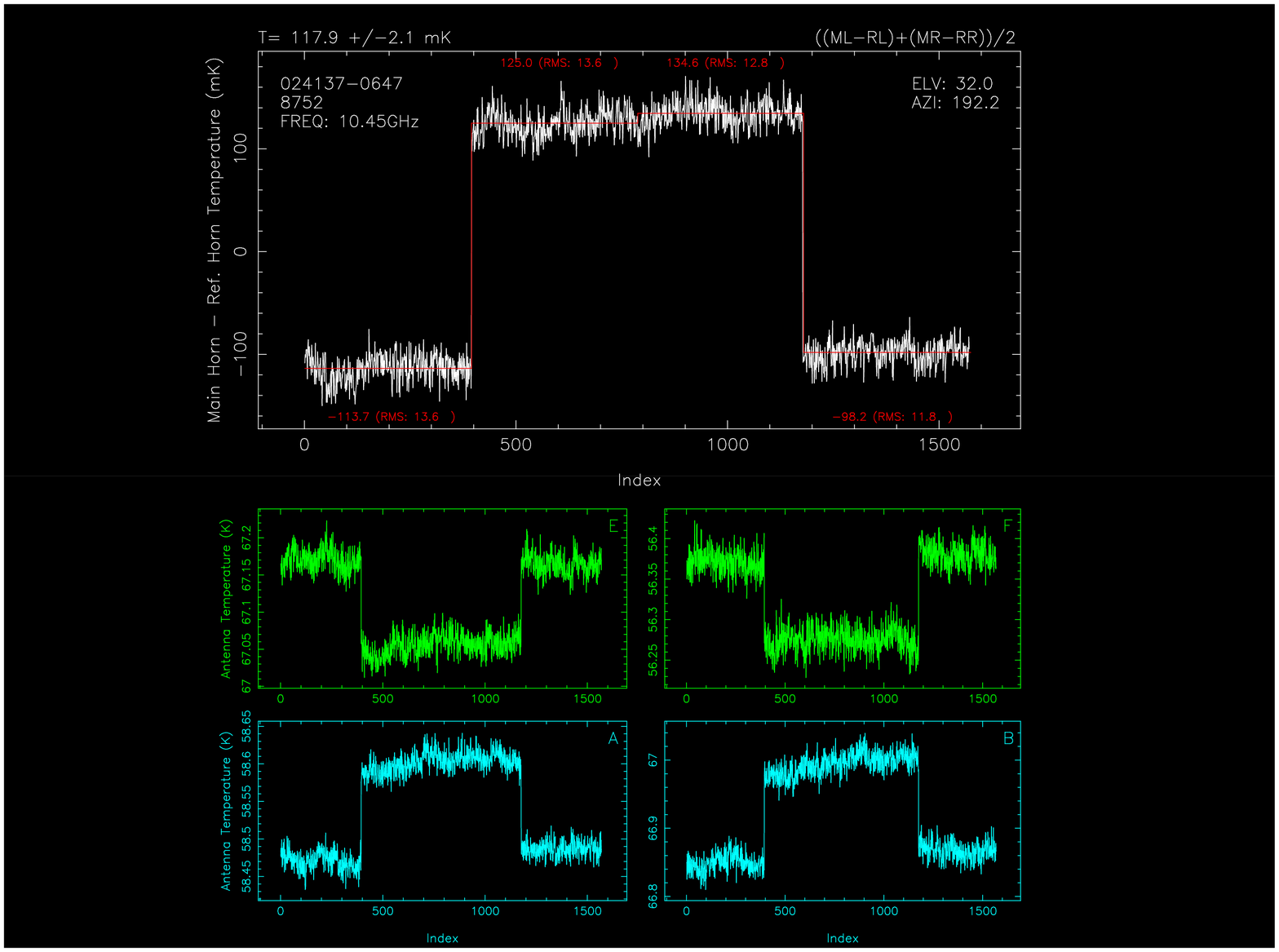}
  \caption{Demonstration of the efficiency of the observing technique
    (upper panel) and a prototype detection profile (lower panel) in
    the case of the 10.45\,GHz receiver. Each receiver has two feeds,
    each of which delivers two channels (LCP and RCP), giving a total
    of four channels. Those are shown in the four lower panels. The
    green colour represents the {\sl reference} horn signal and the
    blue the {\sl main} horn signal. The left-hand side panels are the
    LCP and the right-hand side panels are the RCP. The plot at the
    top of each panel shows the final profile after subtracting the
    signals from each of the two feeds and averaging over LCP and
    RCP. If $MR$ is the RCP of the main horn and $ML$ the LCP in the
    same horn, while $RR$, $RL$ are for the reference horn, the final
    signal is given by $\left[\left(ML-RL\right) +
      \left(MR-RR\right)\right]/2$. It is noteworthy that despite the
    complete absence of even the hint of a source in the individual
    channels (upper panel), after the subtraction a clear case of a
    22-mK signal (roughly 17\,mJy) can be seen.}
  \label{fig:OnOff}
\end{figure}

In order to illustrate the exact observing technique used, we label
the feeds of any of the two receivers as {\sl reference} and {\sl
  main}. Let A be the configuration of having the reference beam
on-source while the main beam is off-source and B the reciprocal
case. The telescope is then slewed in such a way as to complete a
sequence of four sub-scans in a A-B-B-A pattern. Assuming then that
the system temperature is the same in both feeds for any given
sub-scan the differentiation of the two signals should remove any
other contribution than that attributed to the source power. The
efficiency of the method is demonstrated in Fig.~\ref{fig:OnOff}.

Despite its performance , as it is demonstrated in
Fig.~\ref{fig:OnOff}, this technique suffers from two major
disadvantages: (i) it is subject to pointing errors that may result in
power loss. This has been controlled with frequent pointing checks on
strong nearby sources. As shown in Sect.~\ref{subsec:Logistics}
these errors are negligible; (ii) it is subject to cases of confusion
i.e.  cases of sources that contribute power to off-source position
causing a false subtraction result. The solution to that could be
either to observe the target at a different parallactic angle (at which
there would be no confusing source in the off position), or to correct
for it if the power of the confusing source is known. This approach is
discussed in Sect.~\ref{subsec:Logistics}.

\subsection{Logistics}
\label{subsec:Logistics}
\begin{figure}
  \centering
  \includegraphics[width=0.39\textwidth,angle=-90]{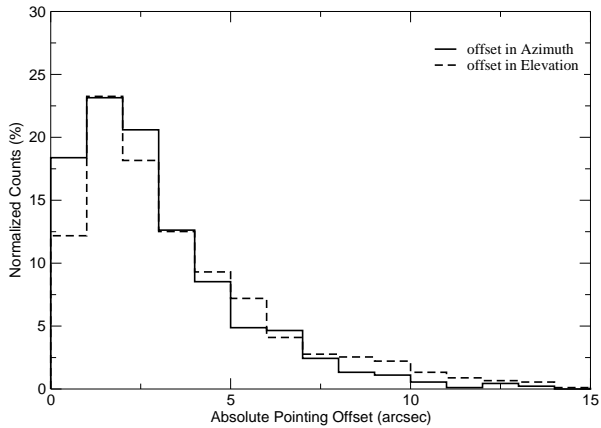}
  \caption{The distribution of the pointing offsets for the case of
    the 10.45\,GHz receiver. The dashed line represents the offsets in
    the elevation direction while the solid one gives those in azimuth. The
    mean offset is around 3\arcsec corresponding to roughly 1\% power
    loss.}
  \label{fig:PoiOff28}
\end{figure}

\paragraph{\bf Thermal noise:}
For both frequencies, the goal of thermal noise
($\sigma_{\mathrm{rms}}$, see also Sect.~\ref{subsec:DataRed}) around
0.2\,mJy (1\,$\sigma$ level) has been set.  Had this been the dominant
noise factor, setting a 5$\sigma$ detection threshold would allow the
detection of sources as weak as 1\,mJy.  The total integration time
for achieving this thermal noise level is 1 and 4 minutes at 4.85\,GHz
and 10.45\,GHz, respectively. This time is the cumulative integration
time for all four sub-scans making up one observing cycle, that is a
scan (see also Sect.~\ref{subsec:Technique}).  However, as shown in
Sect.~\ref{subsec:RepPlotsErr}, the dominant noise factor is the
troposphere rather than thermal noise. It is shown that the practical
limit is of the order of 1.2\,mJy which is judged to be adequate.
\paragraph{\bf Field coverage:}
A rigid constraint is the minimisation of the telescope driving
time. This was achieved by driving the telescope through the field in
a ``zig-zag'' way ({\sl travelling salesman problem}). Each field was
organised in stripes parallel to the right ascension axis and roughly
0.5 degrees across in declination. The sources within such a belt
have, in turn, been organised in dozens in order of monotonous right
ascension change. During an observing session a field would be
targeted within hour angle range from $-3$ to $3$\,hr. This is a
compromise between staying in one field for as long as possible as
well as observing through acceptable airmasses (i.e. not very low
elevations that would result in large opacities).
\paragraph{\bf Pointing offset minimisation and calibration:}
For calibration purposes, one of the standard calibrators shown in
table~\ref{tab:calibrators} was observed at the beginning and the end
of the observation of a field, i.e. roughly every six hours.  Before
the beginning of the field, also the focus of the telescope would be
optimised. Changes in the focal plane within those six hours were
accounted for by interpolation of the sensitivity factor between the
values measured at the beginning and the end of the run. To maintain
low pointing offsets, cross-scans were frequently performed on bright
nearby point sources. On average a pointing check was done every 30
minutes to 1.5 hours. This sustained average pointing offsets of as
low as 3-4\arcsec for the 10.45\,GHz and 7-8\arcsec for the 4.85\,GHz
measurements. These correspond to 4.5\% and 5\% of the $FWHM$ at
10.45\,GHz and 4.85\,GHz respectively and result in a negligible power
loss of the order of 1\%. As an example, Fig.~\ref{fig:PoiOff28} shows
the distribution of pointing offsets for the high frequency
observations.

\subsection{Data Reduction}
\label{subsec:DataRed}
Before any further discussion it must be clarified that despite the
fact that the receivers deliver two circular polarization channels
(namely LCP and RCP, see Sect.~\ref{subsec:System}), the possible
circular polarization has been neglected with the LCP and RCP channels
being averaged (see Fig.~\ref{fig:OnOff} and
appendix~\ref{app:confusion}). This is a reasonable assumption
provided that the average degree of circular polarization of these
sources is expected to be low ($< 0.1\%$,
e.g. \citealt{Weiler1983ApJS,Komesaroff1984MNRAS}).

Figure~\ref{fig:OnOff} illustrates the ``detection pattern''. From
that picture it is clear that a measurement is the difference between
the average antenna temperature of the first and the second sub-scans
($T_{\mathrm{left}}$) as well as that between the third and the fourth
($T_{\mathrm{right}}$). These two differences essentially provide two
independent measurements of the target source. Ideally, the results
should be identical. Differences should be attributed to atmospheric
fluctuations, given that the overlap of the ``off'' and the ``on''
beam is not precisely 100\,\%, as well as confusion (field sources
contributing power in the off-beam position). This effect however,
comprises the most severe uncertainty in the measurement. A detailed
discussion is given in appendix~\ref{sec:ExpectedVsObs}.

Throughout the data reduction process two types of errors are
computed. The first, denoted by $\sigma_{\mathrm{rms}}$, is the
result of the formal error propagation (assuming Gaussian statistics)
of the data scatter around the average (error in mean), is chiefly a
property of the detector and is practically computed by the radiometer
formula.  The second is root mean square (rms) in the antenna
temperature as is measured from the first subtraction (sub-scans 1 and
2) and that from the second subtraction (sub-scans 3 and 4). That is,
$\sigma_{\mathrm{\Delta T}}= |T_{\mathrm{left}}
-T_{\mathrm{right}}|/2$.  Subsequently, the
$max\left(\sigma_{\mathrm{rms}},\sigma_{\mathrm{\Delta T}}\right)$ is
taken as a first estimate of the error in the measurement. In
Sect.~\ref{subsec:RepPlotsErr}, we describe how the final errors
reported in table~\ref{tab:final} have been calculated.

\subsubsection{Corrections}
\label{subsec:corrections}
\begin{figure}
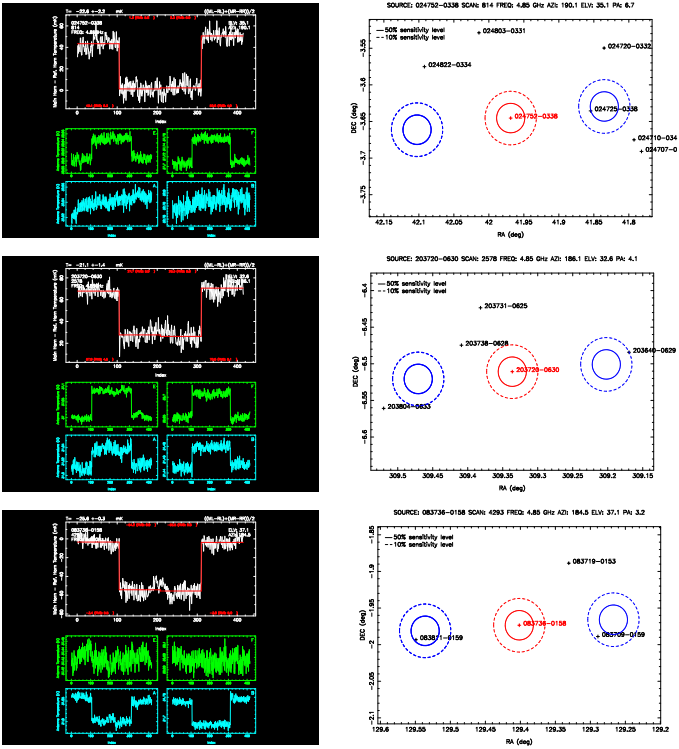

  \centering
  \begin{tabular}{cc}
    \includegraphics[width=0.17\textwidth,angle=-90]{1267fg05.epsi} &
    \includegraphics[width=0.17\textwidth,angle=-90]{1267fg06.epsi}\\
    \includegraphics[width=0.17\textwidth,angle=-90]{1267fg07.epsi} &
    \includegraphics[width=0.17\textwidth,angle=-90]{1267fg08.epsi}\\
    \includegraphics[width=0.17\textwidth,angle=-90]{1267fg09.epsi} &
    \includegraphics[width=0.17\textwidth,angle=-90]{1267fg10.epsi}\\
  \end{tabular}
  \caption{Confusion examples. The left-hand column shows the
    detection profiles whereas the right-hand one shows the NVSS
    environment of each target source. There -- assuming a Gaussian
    beam pattern -- the solid line marks the 50\% beam gain level
    while the dashed one denotes the 10\% level. The red circles
    correspond to the ``on'' positions and blue ones to the ``off''
    positions. The target sources are shown in red and the environment
    NVSS ones in black. The left-hand side plots show the result of
    the differentiation with respect to the strength of the
    ``confusers'' and their position relative to the centre of the
    beam.}
  \label{fig:confusion}
\end{figure}

Each measurement conducted as described earlier is consequently
subjected to a number of corrections:
\paragraph{\bf Opacity correction:}
This process is meant to correct the attenuation of the source signal
due to the terrestrial atmosphere. The computation of the opacity is
done by utilisation of the observed system temperatures.
\paragraph{\bf Elevation dependent gain correction:}
Small scale divergences of the primary reflector's geometry from the
ideal paraboloid lower the sensitivity of the antenna. These
gravitational deformations are a function of elevation with the
consequence of an elevation-dependent antenna gain. The
``elevation-gain'' curve is a second order polynomial of the elevation
and is constructed experimentally by observing bright point-like
sources over the entire elevation range.
\paragraph{\bf Sensitivity correction:}
This process is essentially the translation of the antenna temperature
to Jy. That is done by observing standard calibrators
(table~\ref{tab:calibrators}). Given a source of known flux density
$S_\mathrm{cal}$ [Jy] and measured antenna temperature $T_\mathrm{A}$
[K], the sensitivity factor $\Gamma$ will then be $\Gamma=T_\mathrm{A}
/ S_\mathrm{cal}$. However, the sensitivity factors obtained this way
depend on the quality of axial focusing. This is optimised at the
initialisation of a field observation. Nevertheless, it can change
over the span between two such consecutive optimisations (of the order
of six hours) and particularly when large temperature gradients are
present throughout the telescope structure.  In accounting for that,
the sensitivity factors have been measured both after the first focus
correction (beginning of the observation) and also before the next
focus correction (end of the field observation). For an observing
instant in between, the result of linear interpolation between those
two values has been used.  The flux densities of the calibrators are
taken from \cite{Ott1994}, \cite{Baars1977AnA} and Kraus priv. comm.
It must be noted that apart from NGC\,7027 the sources used as
calibrators are point-like for the beamwidth of Effelsberg
telescope. NGC\,7027 on the other hand, is extended at 10.45\,GHz. At
this frequency its size is roughly 9.0$\times$12.0\arcsec (the
beamwidth at 10.45 GHz is $\sim 67$\arcsec). Nevertheless, the power
loss due to this effect is still less than 1\,\% and therefore, no
beam correction is necessary.
\begin{table}[b]
  \caption{The flux densities and spectral indices of the standard calibrators.}
  \label{tab:calibrators}  
  \centering                    
  \begin{tabular}{l c c c c c c } 
    \hline\hline                 
    Source &$S_{4.85}$ &$\left<S_{4.85}\right>^\dagger$ &$S_{10.45}$ &$\left<S_{10.45}\right>^\dagger$ &$\alpha_{1.4}^{4.85}$   &$\alpha_{4.85}^{10.45}$ \\    
           &(Jy)  &(Jy)      &(Jy) &(Jy)      &                      &   \\    
    \hline                         
    3C\,48     & 5.63 & 5.52  & 2.68  & 2.59   &$-$0.88   &$-$0.98 \\   
    3C\,161    & 6.61 & 6.71  & 2.99  & 3.03   &$-$0.82   &$-$1.03 \\   
    3C\,286    & 7.50 & 7.48  & 4.44  & 4.48   &$-$0.54   &$-$0.67 \\   
    NGC\,7027  & 5.40 & 5.46  & 5.83  & 5.96   &$+$1.12   &$+$0.11 \\   
    \hline                                  
  \end{tabular}
  \begin{list}{}{}
  \item Columns 2 and 4 give the flux densities assumed for these
    sources. The flux densities and the spectral indices in columns 3,
    5 and 6, 7 respectively, are the ones averaged over our
    measurements.
  \item[$^\dagger$] The average has been done over data with $SNR \ge 5\sigma$.
  \end{list}
\end{table}
\paragraph{\bf Confusion:}
A potential limitation for any observation is confusion which has been
well studied since the early days of radio surveys
\citep{Scheuer1957PCPS}. Put simply, it refers to blends of unresolved
sources that build up significant flux densities. Traditionally,
confusion has been treated statistically in terms of expected flux
density per unit area for a given frequency and for modern observing
facilities it often constitutes a factor imposing more severe limits
than the thermal noise itself. In the case of the currently discussed
work, the problem becomes even more severe because of the beam switch
technique used. In this case, any combination of field NVSS sources
can be in the vicinity of the targeted source within the ``on'' or any
of the ``off'' positions. That can severely affect the differentiation
algorithm by contaminating the subtracted signal. Some typical
confusion cases are shown in Fig.~\ref{fig:confusion}.
The confusion status has been monitored for every observed scan on the
basis of the NVSS positions and has been corrected afterwards whenever
possible (see description appendix~\ref{app:confusion}).

\section{Errors}
\label{sec:Errors} 
In general, the requirement of time efficiency can be in conflict with
measurement accuracy by limiting, for instance, the time invested in
calibration. A careful and realistic quantification of the involved
uncertainties is therefore essential.  The following discussion deals
with the system repeatability study which in fact sets the pragmatic
limit to the reliably detectable flux density.

\subsection{System repeatability}
\label{subsec:repeat}
Given the goal of reaching the telescope's theoretically expected
least detectable flux density, it is crucial to estimate the
repeatability of a measurement. Let the term ``observing system''
collectively describe everything but the target source. Hence, it
refers to the combination of the telescope, the thermal noise, the
atmosphere, the confusion etc. An ideal observing system should output
exactly the same result for the flux of a source independently of the
number of repetitions of the measurements, as long as the source
itself is not variable. If we therefore assume that the source is
non-variable, the variance of its measured flux density over several
repetitions can be perceived as system variability caused by any
combination of the possible factors referred to previously. The
estimation of the mean variance of the system as a whole sets the
lower limit in the detectable flux density. Considerable observing
time has been spent in monitoring exactly this property of the system.

A number of sources, hereafter called the ``repeaters'', have been
selected to be observed during every observing run. They have been
chosen to satisfy two conditions:
\begin{enumerate}
\item To be intrinsically non-variable. It is known that sources of
  steep spectrum are unlikely to be intrinsically variable. Therefore,
  a number of sources with spectral index steeper than around $-0.5$
  were chosen.
\item To uniformly cover the whole attempted flux density space. This
  is essential as we expect that the system repeatability, as defined
  above (the rms in the repeatedly measured flux density of an
  intrinsically non-variable source), is a function of the flux
  density of the target.
\end{enumerate}
Roughly 10 sources per field were selected and repeatedly observed at
the beginning of each observing run of the respective field. These
sources are included in table~\ref{tab:repeaters} along with their
average fluxes at 1.4, 4.85\,GHz and 10.45\,GHz as well as their low-
and high-frequency spectral indices. As it is shown there, their flux
densities cover the range up to a few hundred mJy. In order to extend
the flux density range the pointing sources and the main calibrators
were also used in the analysis (see tables~\ref{tab:calibrators} and
\ref{tab:repeaters}).
\begin{table*}
  \caption{
    The sources used for pointing correction and the ``repeaters''.}     
  \label{tab:repeaters}  
  \centering                    
  \begin{tabular}{lrrrrrrrr}
    \hline\hline                 
     Source &\multicolumn{1}{c}{$S_{1.4}$} &\multicolumn{1}{c}{$\left<S_{4.85}\right>^\dagger$} &\multicolumn{1}{c}{{\bf rms}} &\multicolumn{1}{c}{$\left<S_{10.45}\right>^\dagger$} &\multicolumn{1}{c}{{\bf rms}} &\multicolumn{1}{c}{$\alpha_{1.4}^{4.85}$}   &\multicolumn{1}{c}{$\alpha_{1.4}^{10.45}$} \\    
            &\multicolumn{1}{c}{(mJy)} &\multicolumn{1}{c}{(mJy)} &\multicolumn{1}{c}{(mJy)} &\multicolumn{1}{c}{(mJy)} &\multicolumn{1}{c}{(mJy)} & &  \\    
    \hline                     
\multicolumn{8}{c}{Pointing Sources} \\
    024104$-$0815$^\ddagger$ &913    &1\,396 &22 &1\,597 &141  &$+$0.34 &$+$0.28 \\ 
    024137$-$0647 &770    &215    &4  &92     &3    &$-$1.03 &$-$1.06 \\ 
    024240$-$0000 &4\,848 &1\,910 &32 &939    &24   &$-$0.75 &$-$0.82 \\
    085509$-$0715 &1\,157 &419    &9  &198    &4    &$-$0.82 &$-$0.88 \\
    085537$+$0312 &618    &218    &5  &101    &4    &$-$0.84 &$-$0.90 \\
    090225$-$0516 &1\,198 &302    &6  &117    &3    &$-$1.11 &$-$1.16 \\
    144839$+$0018 &1\,652 &568    &7  &252    &9    &$-$0.86 &$-$0.94 \\
    145510$-$0539 &1\,028 &316    &3  &148    &4    &$-$0.95 &$-$0.97 \\
    150334$-$0230 &1\,040 &336    &4  &141    &4    &$-$0.91 &$-$0.99 \\
    203640$-$0629 &1\,045 &980    &28 &796    &24   &$-$0.05 &$-$0.14 \\
    204710$-$0236 &2\,282 &900    &14 &480    &12   &$-$0.75 &$-$0.78 \\
\multicolumn{8}{c}{Repeaters} \\
    024747$+$0131 &277 &75  &2 &31 &2   &$-$1.05 &$-$1.09\\  
    024941$+$0134 &60  &15  &1 &6  &2   &$-$1.13 &$-$1.15\\  
    025020$+$0130 &39  &9   &1 &4  &1   &$-$1.18 &$-$1.18\\  
    025341$+$0100 &562 &138 &3 &52 &2   &$-$1.13 &$-$1.18\\  
    025438$+$0056 &117 &73  &3 &50 &3   &$-$0.38 &$-$0.42\\  
    025515$+$0037$^\ddagger$ &31  &35  &3 &73 &19  &$+$0.10 &$+$0.42\\  
    025613$+$0039 &19  &24  &1 &18 &2   &$ $0.17 &$-$0.06\\  
    025615$+$0057 &17  &10  &1 &7  &1   &$-$0.38 &$-$0.45\\  
    025631$+$0041 &73  &20  &2 &8  &2   &$-$1.06 &$-$1.10\\  
    025800$+$0113 &12  &7   &2 &5  &2   &$-$0.50 &$-$0.43\\  
    025825$+$0103 &36  &9   &1 &4  &1   &$-$1.15 &$-$1.08\\  
    084037$-$0034 &22  &5   &1 &2  &0.4 &$-$1.31 &$-$1.19\\  
    084550$-$0051 &114 &54  &2 &30 &3   &$-$0.60 &$-$0.67\\  
    084601$-$0040 &30  &11  &1 &10 &0.4 &$-$0.81 &$-$0.53\\  
    084709$-$0047 &62  &23  &2 &8  &2   &$-$0.82 &$-$1.02\\  
    084721$-$0025 &72  &19  &2 &8  &1   &$-$1.09 &$-$1.10\\  
    084840$-$0034 &131 &37  &1 &17 &2   &$-$1.03 &$-$1.03\\  
    084950$-$0010 &40  &14  &1 &8  &3   &$-$0.87 &$-$0.82\\  
    085255$-$0023 &32  &13  &1 &7  &1   &$-$0.74 &$-$0.75\\  
    085418$-$0036 &58  &17  &2 &9  &2   &$-$0.98 &$-$0.92\\  
    144043$+$0017 &70  &20  &1 &8  &2   &$-$0.10 &$-$1.10\\  
    144119$+$0025 &84  &24  &2 &11 &2   &$-$1.02 &$-$1.01\\  
    144232$+$0019 &30  &9   &1 &5  &1   &$-$1.02 &$-$0.96\\  
    144615$+$0009 &58  &18  &2 &8  &2   &$-$0.95 &$-$0.98\\  
    145004$+$0024 &51  &13  &1 &5  &1   &$-$1.10 &$-$1.15\\  
    145421$-$0016 &84  &34  &1 &17 &2   &$-$0.73 &$-$0.80\\  
    145430$-$0030 &24  &11  &1 &6  &2   &$-$0.60 &$-$0.70\\  
    145548$-$0037 &67  &36  &2 &19 &2   &$-$0.49 &$-$0.63\\  
    145554$-$0037 &32  &30  &2 &13 &1   &$-$0.07 &$-$0.44\\  
    204952$-$0245 &115 &29  &2 &9  &2   &$-$1.10 &$-$1.27\\  
    205001$-$0249 &261 &95  &2 &46 &2   &$-$0.82 &$-$0.87\\  
    205041$-$0249 &26  &20  &2 &16 &4   &$-$0.20 &$-$0.27\\  
    205240$-$0156 &44  &14  &1 &6  &2   &$-$0.92 &$-$1.01\\  
    205546$-$0204 &94  &24  &1 &10 &2   &$-$1.08 &$-$1.14\\  
    205612$-$0206 &90  &22  &3 &12 &2   &$-$1.15 &$-$1.01\\  
    205616$-$0155 &63  &16  &2 &7  &2   &$-$1.10 &$-$1.13\\  
    \hline                                             
  \end{tabular}                                        
  \begin{list}{}{}
  \item The second column shows their NVSS flux density. The third and
    the fifth columns give the average flux densities at 4.85\,GHz and
    10.45\,GHz measured at Effelsberg, respectively.  Columns 4 and 6
    give the rms scatter in those measurements.
  \item[$^\dagger$] The average values are produced by pure detection
    cases. Only measurements of $SNR \ge 5\sigma$ are accepted.
  \item[$^\ddagger$] Source showing intense variability. Marked with
    red in the repeatability plots at 10.45\,GHz.
  \end{list}
\end{table*}

\subsection{Repeatability plots and error budget}
\label{subsec:RepPlotsErr}
In Sect.~\ref{subsec:DataRed} it was explained that as a first
estimate of the error in a measurement, has been taken the maximum
between the error in the mean after the formal error propagation,
$\sigma_{\mathrm{rms}}$ and the part influenced by the atmospheric
fluctuations and confusion, $\sigma_{\Delta T}$
(i.e. $max\left(\sigma_{\mathrm{rms}},\sigma_{\Delta T}\right)$).  The
former is a parameter of the detector and is not expected to vary
significantly. The latter on the other hand can vary even for the same
target source as a function of the atmospheric conditions and the
geometry of the dual-beam system with respect to the target source and
its NVSS environment (confusion).

\begin{figure}
  \centering
  \includegraphics[width=0.3\textwidth,angle=-90]{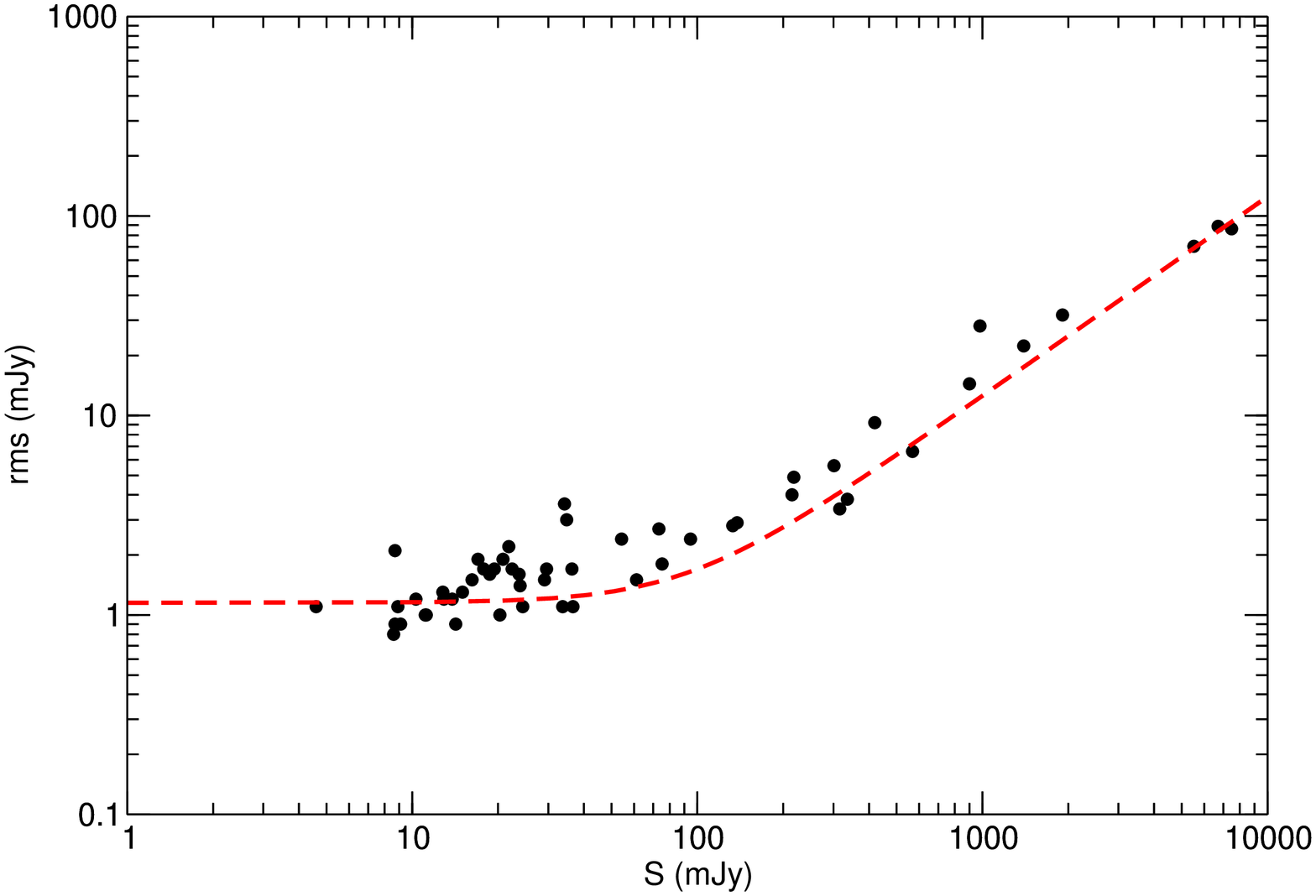}
  \includegraphics[width=0.3\textwidth,angle=-90]{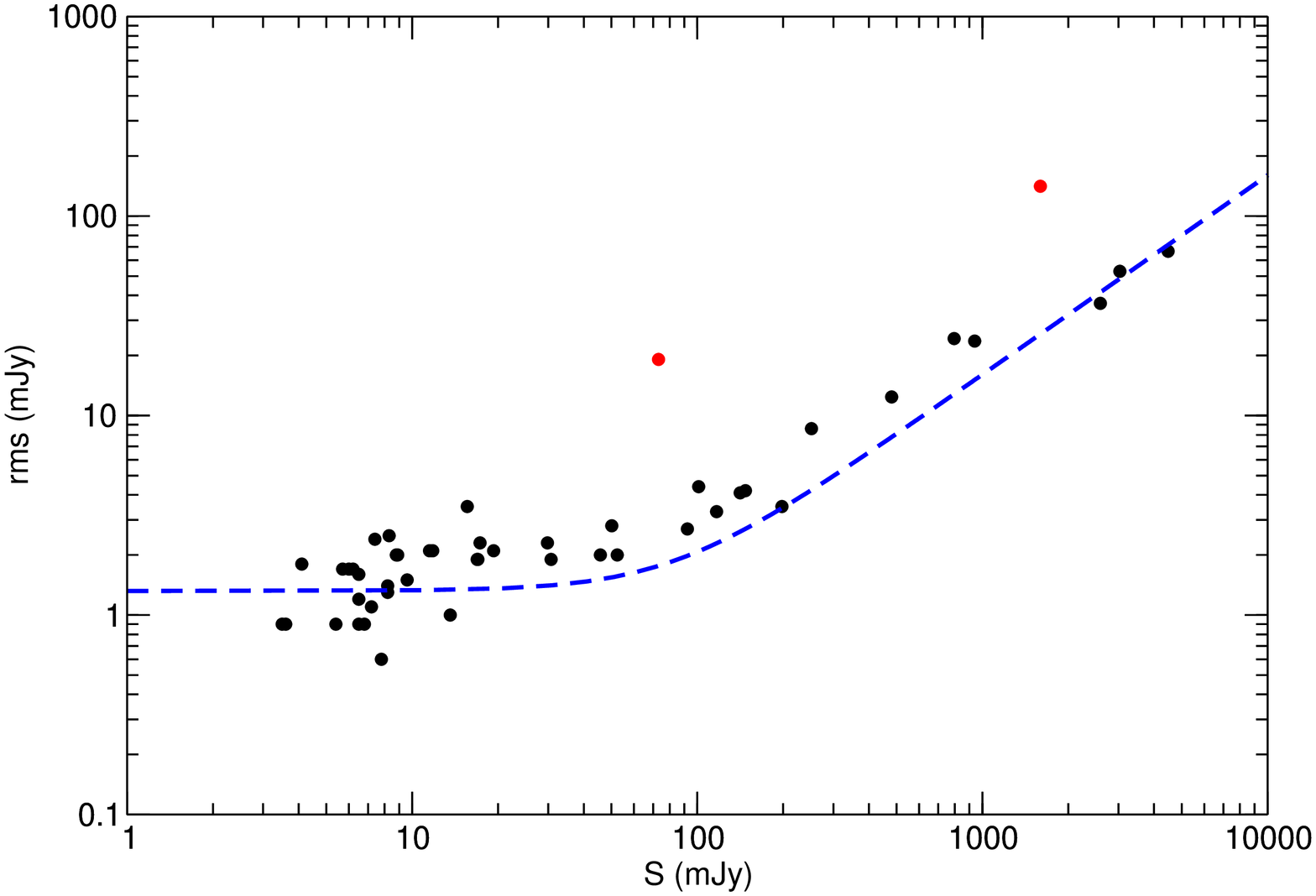}
  \caption{The repeatability plots. The upper plot corresponds to
    4.85\,GHz and the lower one to 10.45\,GHz. The parameters of the
    fitted curves are given in table \ref{tab:fitted}. In the lower
    panel (10.45\,GHz) the red points correspond to sources that are
    known to exhibit variability characteristics and have been
    excluded during the fitting procedure (namely, 025515+0037 and
    024104-0815 in order of flux density).}
  \label{fig:rep}
\end{figure}

A way to statistically quantify the uncertainty in a measurement
including collectively all possible factors of uncertainty, is to
investigate how well the measurement of a target source, assumed
intrinsically non-variable, repeats over several observations (see
Sect.~\ref{subsec:repeat}).  For a given frequency, the measure of the
system repeatability is the rms in the average flux for every repeater
as a function of its average flux density,
$\sigma_{\mathrm{rep}}\left(S \right)$.  The associated plots and are
shown in Fig.~\ref{fig:rep}.

The rms flux density $\sigma_{\mathrm{rep}} \left(S \right)$, can be
written as a function of the mean flux density $S$, being the
Pythagorean sum of (i) the flux density independent term, $\sigma_{0}$
and (ii) the flux density dependent term, $m\cdot S$. In particular,
it is described by:
\begin{equation}
  \label{eq:error}
  \sigma_{\mathrm{rep}} \left(S \right) = \sqrt{ \sigma_{0}^{2} +  \left(m\cdot S \right)^{2}  }
\end{equation}

Fitting this function to the 4.85\,GHz and 10.45\,GHz measurements,
has resulted in the parameters in table~\ref{tab:fitted}.
\begin{table}[b]
  \caption{The fitted parameters for the repeatability curves.}
  \label{tab:fitted}  
  \centering                    
  \begin{tabular}{ccccc} 
    \hline\hline
    Frequency  &$\sigma_{0}$ &Error  &$m$  &Error  \\    
    (GHz) &(mJy) &(mJy)  &(\%) &(\%)  \\    
    \hline                         
    4.85          &1.2  &0.2  &1.3 &0.02   \\
    10.45         &1.3  &0.1  &1.6 &0.04   \\
    \hline
  \end{tabular}
  \begin{list}{}{}
  \item The $\sigma_{0}$ parameter is the one determining the least
    detectable flux.
  \end{list}
\end{table}
\noindent
From those fits one can readily estimate the minimum detectable flux
density at each frequency. Setting the detection threshold at
$5\,\sigma$, the smallest detectable flux density is roughly
6\,mJy. In appendix~\ref{sec:ExpectedVsObs} we discuss the comparison
of the fitted values of $\sigma_{0}$ with the error of an individual
measurement and give a quasi-empirical interpretation of the measured
parameters.  From the discussion previously and in
Sect.~\ref{subsec:DataRed} the most reasonable (and rather
conservative) estimate of the errors, would be:
\begin{equation}
  \label{eq:finalerror}
  err=max\left(\sigma_{\mathrm{rms}},\sigma_{\mathrm{\Delta T}},\sigma_{\mathrm{rep}}\right) 
\end{equation}
This definition is used to derive the final errors reported in
table~\ref{tab:final}.

\subsection{Confusion flavors}
\label{subsec:confflavors}
\begin{figure}
  \centering
  \includegraphics[width=0.33\textwidth,angle=-90]{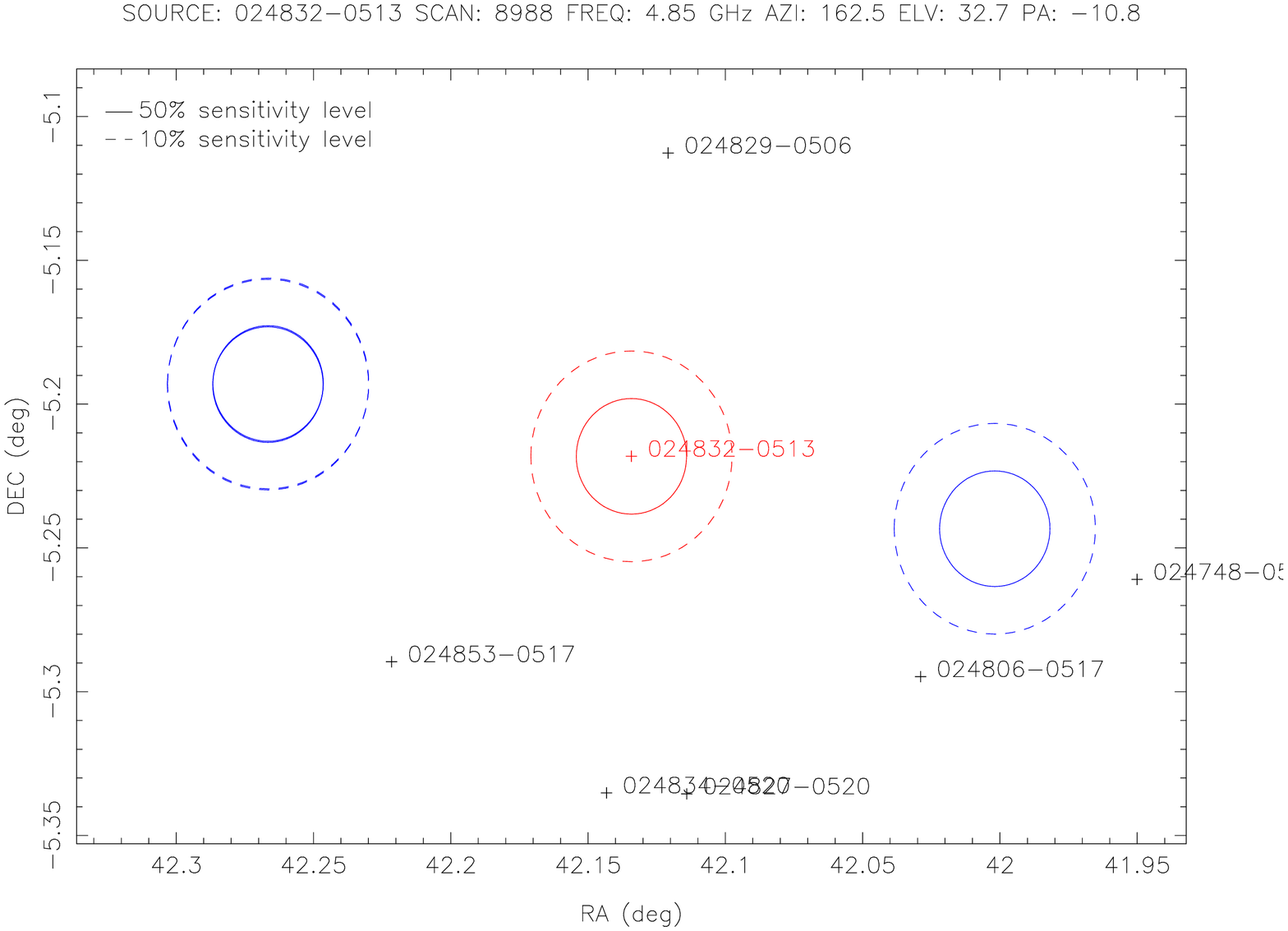} 
  \includegraphics[width=0.33\textwidth,angle=-90]{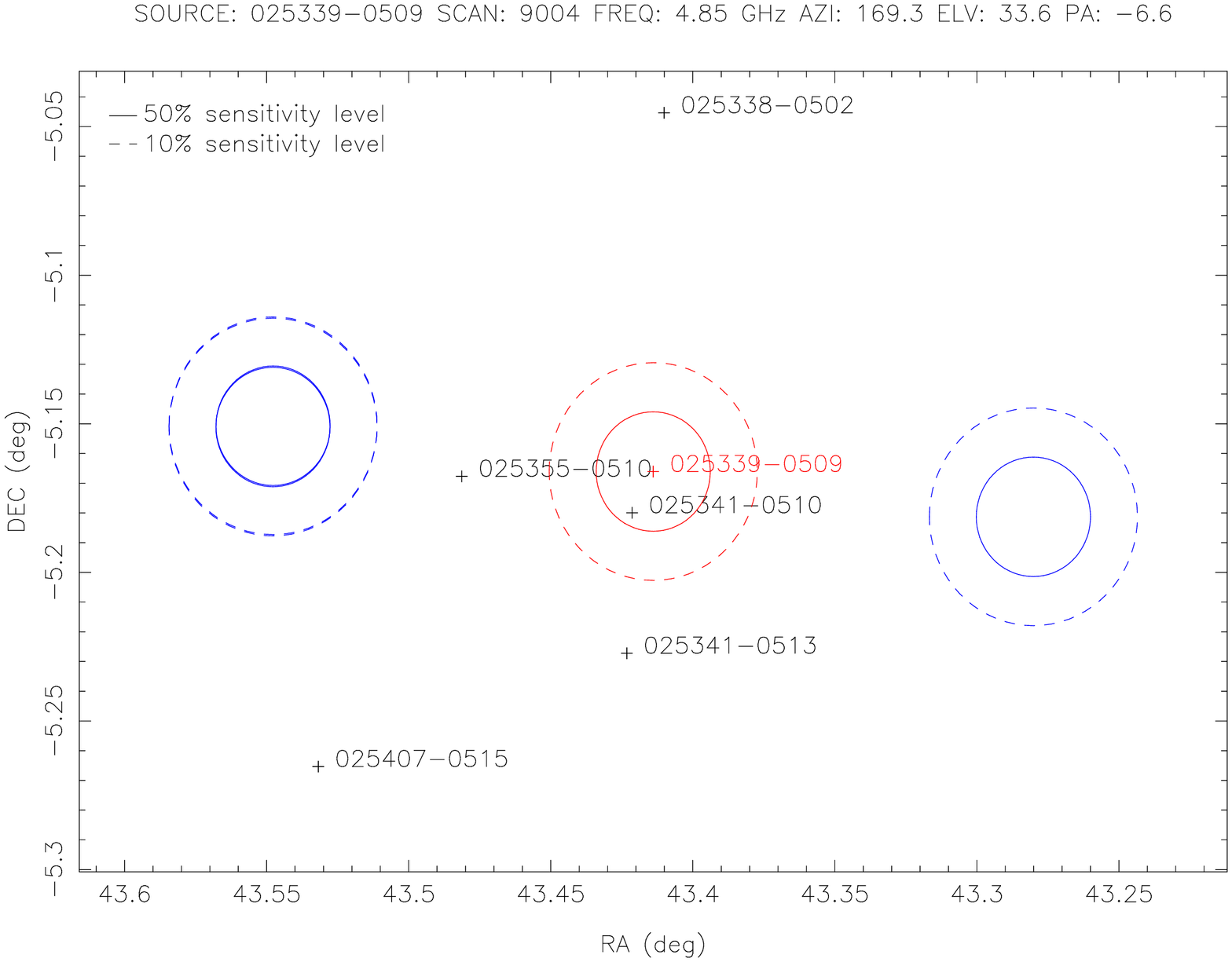} 
  \includegraphics[width=0.33\textwidth,angle=-90]{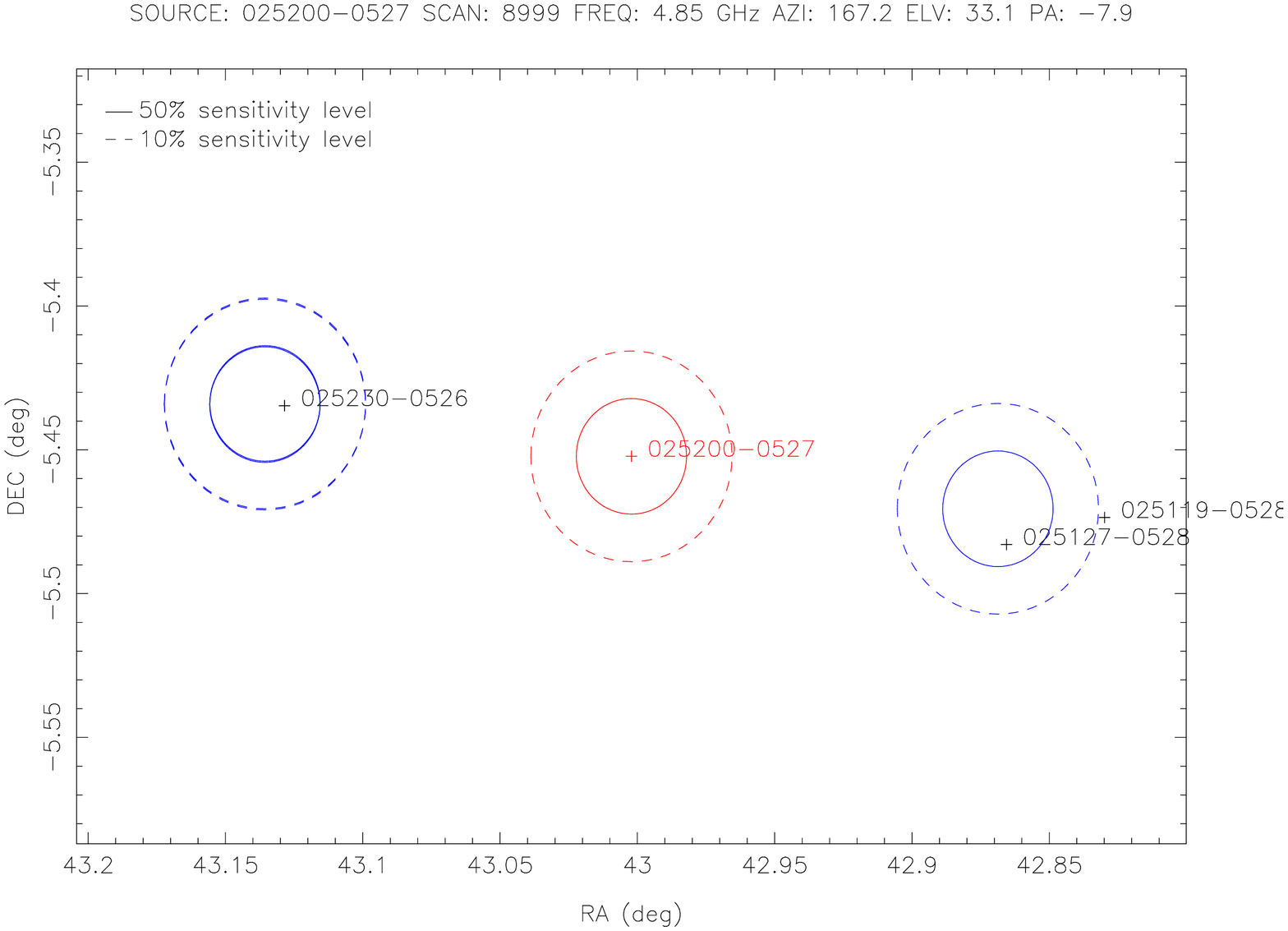} 
  \caption{Confusion flavors. From top to bottom: a {\sl clean}, a
    {\sl cluster} and a {\sl confusion} case. The notation is
    identical to that in Fig.~\ref{fig:confusion}.}
  \label{fig:confflvors}
\end{figure}

Depending on the configuration of the dual-beam system in the sky
relative to the target source and the instantaneous spatial
distribution of NVSS field sources, a given scan can show different
confusion ``flavors''. On this basis, there can be three scan classes
discriminated by their confusion status at a given observing
frequency:
\begin{enumerate}
\item {\sl Clean} scans. Those are the measurements during which there
  were no contaminating sources within any of the beam positions (see
  top panel in Fig.~\ref{fig:confflvors}). For these cases, further
  action need not be taken.
\item {\sl Clustered} scans. These are scans on sources that are
  accompanied by neighboring sources within a radius smaller than the
  associated beam-width (hence the term {\sl cluster}). These sources
  cannot be discriminated (see middle panel in
  Fig.~\ref{fig:confflvors}) and reliable measurement is
  impossible. For a given frequency only an instrument of larger
  aperture could resolve them (e.g. interferometer). For this reason,
  these scans are absent from the discussions in the current paper.
\item {\sl Confused} scans. This refers to the case of having any
  combination of field sources within any beam (see lower panel in
  Fig.~\ref{fig:confflvors}). For these cases one must either conduct
  a measurement at different parallactic angle such that the confusing
  source will not coincide with an ``off'' position, or reconstruct
  their flux from the exact knowledge of the flux of the ``confusers''
  (see appendix~\ref{app:confusion}).
\end{enumerate}
It is important to underline that this effect refers to confusion from
field NVSS sources alone and not to blends of unresolved background
radio sources which may contribute significant flux.

In table~\ref{tab:confflavors} we show the detected confusion flavors
for each field and frequency. From this table it is readily noticeable
that the confusion becomes less important with increasing frequency.
For instance, the fraction of sources that suffer neither from
clustering nor from confusion effects increases from 59\% at 4.85\,GHz
to 92\% at 10.45\,GHz. This is easily interpretable in terms of
smaller beam-width (67\arcsec as opposed to 146\arcsec at
4.85\,GHz). In fact, considering that the majority of sources show
steep radio spectra (see Sect.~\ref{subsec:si}), it is expected that
in practice significantly fewer sources will suffer from confusion
simply because their field sources are too weak already at 4.85\,GHz
and 10.45\,GHz. It is important to state that in the following studies
we consider only a sub-sample of 3\,434 sources which are either {\sl
  clean} or have been {\sl de-confused}.
\begin{table}[h]
  \caption{The frequencies of confusion flavors of the observed 
    scans (measurements) for each field and observing frequency.}
  \label{tab:confflavors}  
  \centering
  \begin{tabular}{ccrrr} 
    \hline\hline                 
    Field  &{\sl Clean} &\multicolumn{1}{c}{\sl Cluster$^\star$} &\multicolumn{1}{c}{\sl Confused}&\multicolumn{1}{c}{\sl de-confused} \\    
           &(\%) &\multicolumn{1}{c}{(\%)} &\multicolumn{1}{c}{(\%)} &\multicolumn{1}{c}{(\%)} \\    
    \hline    
    \\
         &\multicolumn{4}{c}{4.85\,GHz ($FWHM\approx145\arcsec$)} \\
    \cline{2-5}                   
    02$^\mathrm{h}$     &59  &18  &23 &4 (17)\\
    08$^\mathrm{h}$     &57  &18  &25 &9 (36)\\
    14$^\mathrm{h}$     &60  &20  &20 &4 (20)\\
    20$^\mathrm{h}$     &59  &19  &22 &6 (27) \\
    Average      &59  &19  &22 &6 (27) \\ 
    \\
         &\multicolumn{4}{c}{10.45\,GHz ($FWHM\approx67\arcsec$)} \\
    \cline{2-5}                   
    02$^\mathrm{h}$     &92  &1  &7  &0  (0) \\
    08$^\mathrm{h}$     &93  &2  &5  &0  (0) \\
    14$^\mathrm{h}$     &90  &1  &9  &2 (22)\\
    20$^\mathrm{h}$     &92  &2  &6  &1 (17) \\
  Average       &92  &$\sim 1$  &$\sim 7$  &$\sim 2$ (20)\\
    \hline                                  
  \end{tabular} 
  \begin{list}{}{}
  \item It must be made clear that this result is based solely on the
    position of the dual-beam system in the sky and the spatial
    distribution of the field NVSS radio sources. In column 5 we give
    the fraction of the total number of scans that have been
    de-confused and in parenthesis the same number but as a fraction
    of the confused scans which is given in column 4.
  \item[$^\star$] Here the term {\sl cluster} is meant to represent
    both, the cases of pure clustering flavour and those that are
    clustered and confused simultaneously.
  \end{list}
\end{table}

\section{Results} 
\label{sec:results}

\subsection{Detection Rates} 
\label{bozomath}
The essence of our task is identifying the detection rates at each
observing frequency. Assuming that the detectability of a target
source is solely due to its spectral behavior, the detection rates can
reveal the subset of sources that exhibit flat or inverted spectra
which can adversely affect CBI data (i.e. $\alpha \ge -0.5$ with $S
\propto \nu^{\alpha}$). The current sub-section deals with this
problem. That is, essentially counting the sources that have been
detected at each frequency.  For both frequencies the detection
threshold has been set to $5\,\sigma$, with $\sigma$ being the error
in the individual measurement as defined by Eq.~\ref{eq:finalerror}.

A supervisory way to describe the detection rates is using the {\sl
  2-bit binary detection descriptor} as in table~\ref{tab:rates}. That
is, a two-bit binary in which the left-hand side bit describes the
detection at the low frequency and the one on the right-hand side that
at the high frequency with ``0'' denoting a non-detection and ``1''
denoting a detection. From all the sources in the sample we have
selected only those that are either {\sl clean} at 4.85\,GHz or have
been de-confused as described in appendix~\ref{app:confusion}. Those
sources must then also be {\sl clean} at 10.45 where the beam-width is
significantly smaller.
\begin{table}[h]
  \caption{The detection rates.}
  \label{tab:rates}  
  \centering                    
  \begin{tabular}{cccccc} 
    \hline\hline                 
    Field  &sample$^{\star}$ &00     &10   &11   &01   \\    
           &                &(\%)   &(\%) &(\%) &(\%) \\    
    \hline                         
    02$^\mathrm{h}$          &914 &70.5 &16.0 &12.7 &0.8\\
    08$^\mathrm{h}$          &692 &66.8 &19.2 &12.4 &1.6\\
    14$^\mathrm{h}$          &923 &70.7 &18.2 &10.6 &0.5\\
    20$^\mathrm{h}$          &905 &67.8 &19.6 &11.3 &1.3\\
    \\
    Total                   &3434 &69.0 &18.3 &11.7 &1.0 \\
    \hline                                  
  \end{tabular}
  \begin{list}{}{}
  \item The detection threshold for either frequency has been set to
    5\,$\sigma$ with sigma being the error in the individual
    measurement as given by Eq.~\ref{eq:finalerror}.
  \item[$^\star$] The sample includes measurements that are {\sl
      clean} or {\sl de-confused}. The de-confusion includes also the
    rare cases of having the source observed at different parallactic
    angle.
  \end{list}
\end{table}

\subsection{The measured flux densities}
\label{sect:measuredfluxdensities}
In table~\ref{tab:final}, available at the CDS, we summarise the
acquired Effelsberg measurements along with the computed spectral
indices for each source. For the construction of this table, only {\sl
  clean} or {\sl de-confused} cases have been considered.  In that
table, Column 1 lists the name of the source, Columns 2 and 3 give the
NVSS flux density and its error respectively, Columns 4 and 5 give the
flux density at 4.85 GHz and its error respectively. Columns 6 and 7
list the flux density at 10.45 GHz and its error.  The 4.85\,GHz and
10.45\,GHz have been measured with 100\,m radio telescope in
Effelsberg. Columns 8 and 9 give the {\sl low frequency} spectral
index $\alpha_{1.4}^{4.85}$ between 1.4\,GHz and 4.85\,GHz and its
error. Similarly, Columns 10 and 11 give the {\sl high frequency}
spectral index $\alpha_{4.85}^{10.45}$ between 4.85\,GHz and
10.45\,GHz and its error. Finally, Columns 12 and 13, give the
least-square fit spectral index $\alpha$, from the 1.4\,GHz (NVSS),
4.85\,GHz and 10.45\,GHz data points and its error.

As mentioned earlier, a measurement is regarded to be a detection only
if it is characterized by $SNR \ge 5$. Whenever this is not the case
an upper limit of $5 \sigma$ is put, with sigma being the error
computed as described in Sect.~\ref{subsec:DataRed}. In such cases,
the associated spectral indices are not quoted in the table.

According to our discussion in Sect.~\ref{subsec:RepPlotsErr}, the
errors quoted in table~\ref{tab:final} can not be $< 1.2$ or $< 1.3$
at 4.85 or 10.45\,GHz, respectively. Yet, there exist entries that the
given error is less than those limiting values. The reason for this is
that in the rare cases that more than one measurements of the same
source are available a weighted averaged is computed to be the flux
density if the source. Then the associated error may appear smaller
that the values of 1.2 and 1.3.

In that table are included all the measurements that are characterized
as clean or they have been de-confused. Clustered sources or cases
suffering from confusion are not included. It must be noted that
concerning the CMB experiments these cases still provide useful
information. Typically, they are characterized by lower angular
resolutions and hence clustered sources can be treated as individual
objects.  All in all, the sources included there amount to about 57\%
of the whole sample amounting to 3\,434 entries.
\begin{table*}[htb]
  \caption{The Effelsberg measured flux densities along with the NVSS ones and the computed spectral indices (it is assumed that $S\sim \nu^{\alpha}$).}
\label{tab:final}   
\centering                    
  \begin{tabular}{l r r r r r r r c r c r c} 
    \hline\hline
\multicolumn{1}{c}{Source} &\multicolumn{1}{c}{$S_{1.4}$} &\multicolumn{1}{c}{err} &\multicolumn{1}{c}{$S_{4.85}$} &\multicolumn{1}{c}{err} &\multicolumn{1}{c}{$S_{10.45}$} &\multicolumn{1}{c}{err}  &\multicolumn{1}{c}{$\alpha_{1.4}^{4.85}$} &err &\multicolumn{1}{c}{$\alpha_{4.85}^{10.45}$} &err &\multicolumn{1}{c}{$\alpha$} &err\\
&\multicolumn{1}{c}{(mJy)} &\multicolumn{1}{c}{(mJy)} &\multicolumn{1}{c}{(mJy)} &\multicolumn{1}{c}{(mJy)} &\multicolumn{1}{c}{(mJy)} &\multicolumn{1}{c}{(mJy)} & & & &  &  &\\
    \hline                         
023958+0137 &4.0 &0.5 &$<$12.5 &  &$<$8.0 &  & &  &  &  & &  \\
024013+0127 &9.4 &0.6 &$<$6.0 &  &$<$6.5 &  & &  &  &  & &  \\
024031+0140 &3.6 &0.5 &$<$18.5 &  &$<$9.5 &  & &  &  &  & &  \\
024057+0129 &7.3 &0.6 &$<$6.0 &  &$<$6.5 &  & &  &  &  & &  \\
024058+0133 &7.5 &0.5 &$<$7.5 &  &$<$12.0 &  & &  &  &  & &  \\
024119+0126 &7.8 &0.6 &9.3 &1.2 &$<$16.0 &  &$+$0.14 &0.12 &  &  & &  \\
024124+0154 &49.0 &1.5 &15.6 &2.0 &$<$14.5 &  &$-$0.92 &0.11 &  &  & &  \\
024234+0135 &58.2 &1.8 &35.7 &1.3 &24.9 &1.4 &$-$0.39 &0.04 &$-$0.47 &0.09 &$-$0.41 &0.02  \\
024402+0140 &4.2 &0.5 &$<$19.5 &  &$<$6.5 &  & &  &  &  & &  \\
024456+0141 &12.3 &0.6 &$<$7.5 &  &$<$8.0 &  & &  &  &  & &  \\
    \hline                         
 \end{tabular}
  \begin{list}{}{}
  \item Cases of SNR$< 5$ at either of 4.85 or 10.45 GHz are noted
    with upper limits of $5\, \sigma$. The source name is marked with
    $^{\star}$ or $^{\dagger}$ in case de-confusion correction has
    been applied at 4.85 or 10.45 GHz, respectively. The complete
    table is available in electronic form at the CDS via anonymous ftp
    to cdsarc.u-strasbg.fr (130.79.128.5) or via
    http://cdsweb.u-strasbg.fr/cgi-bin/qcat?J/A+A/
  \end{list}
\end{table*}

\subsection{Spectral Indices} 
\label{subsec:si}
The motivation for the current program has been, as discussed earlier,
the estimation of the extrapolated flux to be expected at higher
frequency bands performed on the basis of the three-point spectral
index. Here we summarize the findings of the spectral indices
study. Hereafter, it is assumed that $S\propto {\nu}^\alpha$.

\begin{figure}
  \centering
  \includegraphics[width=0.4\textwidth,angle=-90]{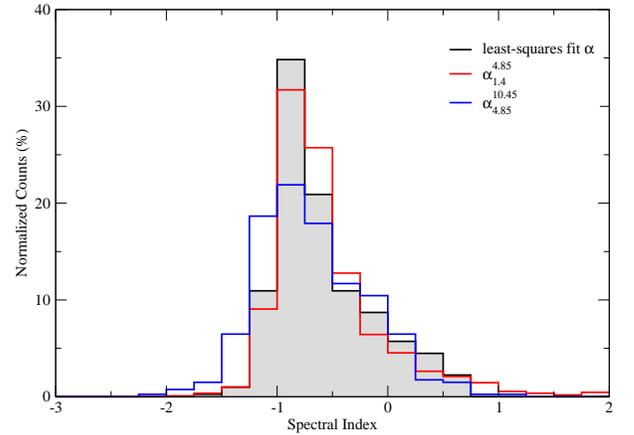}
  \caption{The normalized spectral index distributions. The grey area
    histogram shows the distribution of the three-point
    least-square-fit spectral index, $\alpha$, for sources that have
    been detected at both frequencies with sigma-to-noise of at least
    5 (402 sources, table~ \ref{tab:rates}). The same sub-sample is
    used for the blue line distribution which denotes that of the
    "high" frequency spectral index,
    $\alpha_{4.85}^{10.45}$. Finally, the red line shows the
    distribution of the "low" frequency index, $\alpha_{1.4}^{4.85}$,
    for a number of 1104 sources detected at 4.85 GHz (see table~
    \ref{tab:rates}). The mean values of $\alpha$,
    $\alpha_{1.4}^{4.85}$ and $\alpha_{4.85}^{10.45}$ are $-0.59$,
    $-0.54$ and $-0.69$, respectively.}
  \label{fig:si55.11}
\end{figure}

To begin with, Fig.~\ref{fig:si55.11} shows the spectral index
distributions for the spectral indices in table~\ref{tab:final}. In
particular, the distributions of $\alpha_{1.4}^{4.85}$,
$\alpha_{4.85}^{10.45}$ and least-squares fit three-point $\alpha$ are
shown. All three of those are constructed only with 5\,$\sigma$ data.
For computing the three-point spectral index, an implementation of the
nonlinear least-squares (NLS) Marquardt-Levenberg algorithm
\citep{Marquardt196306} was used. That imposes natural weighting
(i.e. $1/{\sigma}^{2}$).

The median spectral index $\alpha$ is around $-0.71$ whereas the
average value is roughly $-0.59$ indicating the skewness of the
distribution. On the other hand, $\alpha_{1.4}^{4.85}$ shows a median
value of roughly $-0.69$ whereas $\alpha_{4.85}^{10.45}$ has also a
median of $-0.75$. Of the 402 sources detected at both frequencies,
136 (34\%) appear with a spectral index $\alpha \ge -0.5$ which
implies that the majority of the sources appear to have steep spectrum
($\alpha \le -0.5$). Moreover, it nicely explains the large percentage
of sources $\approx 87\%$ with 2-bit binary detection descriptor of 00
or 10. What is important about this population is that these are the
sources that need not be ``vetoed out'' during the CMB data analysis
since they are not bright enough to contribute detectable flux at the
frequencies near 30\,GHz at which experiments like CBI operate.

All the measurements have been conducted in an approximately
quasi-simultaneous way. The coherence time varies between hours to
days. It is therefore important to contemplate on how the lack of
simultaneity influences the results. Provided that most of the sources
follow a steep spectrum trend and steep spectrum sources are not
expected to vary significantly, it is reasonable to assume that
statistically it will be insignificant.

\section{Conclusions}

\begin{enumerate}
\item The applied observing technique has been chosen to be efficient
  in terms of time. Its combination with the beam-switch allows a
  remarkably efficient removal of linear atmospheric effects. However,
  it suffers from ``analytic'' confusion (caused by sources of
  positions known from other surveys) as expected.  Nevertheless, the
  confusion effect decreases fast with frequency (from $\sim\,22 \%$
  to $\sim 6\,\%$ between 4.85\,GHz and 10.45\,GHz) thanks to the
  increase of the telescope angular resolution.  Accounting for the
  shape of the radio spectrum would imply a further decrease in the
  number of sources that can actually cause confusion.
\item We show that for both the 4.85\,GHz and the 10.45\,GHz
  observations the dominant factor in the smallest reliably
  ($\mathrm{SNR}\ge 5$) detectable flux density has been the
  tropospheric turbulence. In Appendix~\ref{sec:ExpectedVsObs} we show
  that the tropospheric factor is of the order of 0.9\,mJy and
  1.3\,mJy for the 4.85\,GHz and the 10.45\,GHz observations,
  respectively. On the other hand, while the second most important
  factor for the low frequency is the confusion caused by blends of
  unresolved sources (see Section~\ref{subsec:RepPlotsErr} and
  Table~\ref{tab:fitted}), for the higher frequency thermal
  ``receiver'' noise dominates. The confusion in the latter case drops
  dramatically by an order of magnitude to 0.08\,mJy due to the
  smaller beamwidth and the presumed spectral behaviour of radio
  sources. From this discussion it is clear that the major limiting
  factor has been the troposphere itself setting a physical limitation
  in the least detectable flux density. That appears to be between
  $5\times 1.2=6$ and $5\times 1.3=6.5$ for the 4.85\,GHz and the
  10.45\,GHz respectively.
\item The agreement between the interpretation/formulation of the
  errors described in Appendix~\ref{sec:ExpectedVsObs} and the
  observed ones from the study of the ``repeaters'' is noteworthy.
\item In Appendix~\ref{app:confusion} an algorithm for achieving
  ``de-confusion'' is presented. That is, reconstructing a source
  antenna temperature on the basis of some elementary
  presumptions. The algorithm has been successfully used in 6\,\% of
  the cases in the current study and can be easily generalised in
  projects demanding automation.
\item In Appendix~\ref{sec:cbionon} we present an algorithm that is
  responsible for the ``quality check'' of every observation (a
  ``scan''). Incorporating a number of tests can be used for
  automatically detecting cases of bad quality data and can be
  generalised to be used in a ``blind'' mode.
\end{enumerate}

\begin{acknowledgements}
  The authors would like to thank the anonymous referee for the
  comments that significantly improved the content of the
  manuscript. Furthermore, we want to thank the internal referee Dr
  D. Graham also for his comments and suggestions.  We would like to
  acknowledge the help of Dr I. Agudo, Mrs S. Bernhart, Dr
  V. M. C. Impellizzeri and Dr R. Reeves and all the operators at the
  100\,m telescope for their help with the observations. The author
  was mostly financed by EC funding under the contract
  HPRN-CT-2002-00321 (ENIGMA) and completed this work as member of the
  International Max Planck Research School (IMPRS) for Radio and
  Infrared Astronomy. All the results presented here have been based
  on observations with the 100\,m telescope of the MPIfR
  (Max-Planck-Institut f\"ur Radioastronomie).
\end{acknowledgements}

\bibliographystyle{aa} 
\bibliography{/Users/mangel/my_research/Literature/MyBIB/References.bib} 

\begin{thebibliography}{18}
\expandafter\ifx\csname natexlab\endcsname\relax\def\natexlab#1{#1}\fi

\bibitem[{{Baars} {et~al.}(1977){Baars}, {Genzel}, {Pauliny-Toth}, \&
  {Witzel}}]{Baars1977AnA}
{Baars}, J.~W.~M., {Genzel}, R., {Pauliny-Toth}, I.~I.~K., \& {Witzel}, A.
  1977, \aap, 61, 99

\bibitem[{{Condon} {et~al.}(1989){Condon}, {Broderick}, \&
  {Seielstad}}]{Condon1989AJ}
{Condon}, J.~J., {Broderick}, J.~J., \& {Seielstad}, G.~A. 1989, \aj, 97, 1064

\bibitem[{{Condon} {et~al.}(1998){Condon}, {Cotton}, {Greisen}, {Yin},
  {Perley}, {Taylor}, \& {Broderick}}]{Condon1998AJ}
{Condon}, J.~J., {Cotton}, W.~D., {Greisen}, E.~W., {et~al.} 1998, \aj, 115,
  1693

\bibitem[{{Komesaroff} {et~al.}(1984){Komesaroff}, {Roberts}, {Milne},
  {Rayner}, \& {Cooke}}]{Komesaroff1984MNRAS}
{Komesaroff}, M.~M., {Roberts}, J.~A., {Milne}, D.~K., {Rayner}, P.~T., \&
  {Cooke}, D.~J. 1984, \mnras, 208, 409

\bibitem[{{Marquardt}(1963)}]{Marquardt196306}
{Marquardt}, D.~W. 1963, Journal of the Society for Industrial and Applied
  Mathematics, 11, 431

\bibitem[{{Mason} {et~al.}(2003){Mason}, {Pearson}, {Readhead}, {Shepherd},
  {Sievers}, {Udomprasert}, {Cartwright}, {Farmer}, {Padin}, {Myers}, {Bond},
  {Contaldi}, {Pen}, {Prunet}, {Pogosyan}, {Carlstrom}, {Kovac}, {Leitch},
  {Pryke}, {Halverson}, {Holzapfel}, {Altamirano}, {Bronfman}, {Casassus},
  {May}, \& {Joy}}]{Mason2003ApJ}
{Mason}, B.~S., {Pearson}, T.~J., {Readhead}, A.~C.~S., {et~al.} 2003, \apj,
  591, 540

\bibitem[{{Ott} {et~al.}(1994){Ott}, {Witzel}, {Quirrenbach}, {Krichbaum},
  {Standke}, {Schalinski}, \& {Hummel}}]{Ott1994}
{Ott}, M., {Witzel}, A., {Quirrenbach}, A., {et~al.} 1994, \aap, 284, 331

\bibitem[{{Padin} {et~al.}(2001){Padin}, {Cartwright}, {Mason}, {Pearson},
  {Readhead}, {Shepherd}, {Sievers}, {Udomprasert}, {Holzapfel}, {Myers},
  {Carlstrom}, {Leitch}, {Joy}, {Bronfman}, \& {May}}]{Padin2001ApJ}
{Padin}, S., {Cartwright}, J.~K., {Mason}, B.~S., {et~al.} 2001, \apjl, 549, L1

\bibitem[{{Padin} {et~al.}(2002){Padin}, {Shepherd}, {Cartwright}, {Keeney},
  {Mason}, {Pearson}, {Readhead}, {Schaal}, {Sievers}, {Udomprasert},
  {Yamasaki}, {Holzapfel}, {Carlstrom}, {Joy}, {Myers}, \&
  {Otarola}}]{Padin2002PASP}
{Padin}, S., {Shepherd}, M.~C., {Cartwright}, J.~K., {et~al.} 2002, \pasp, 114,
  83

\bibitem[{{Pearson} {et~al.}(2003){Pearson}, {Mason}, {Readhead}, {Shepherd},
  {Sievers}, {Udomprasert}, {Cartwright}, {Farmer}, {Padin}, {Myers}, {Bond},
  {Contaldi}, {Pen}, {Prunet}, {Pogosyan}, {Carlstrom}, {Kovac}, {Leitch},
  {Pryke}, {Halverson}, {Holzapfel}, {Altamirano}, {Bronfman}, {Casassus},
  {May}, \& {Joy}}]{Pearson2003ApJ}
{Pearson}, T.~J., {Mason}, B.~S., {Readhead}, A.~C.~S., {et~al.} 2003, \apj,
  591, 556

\bibitem[{{Readhead} {et~al.}(2004){Readhead}, {Mason}, {Contaldi}, {Pearson},
  {Bond}, {Myers}, {Padin}, {Sievers}, {Cartwright}, {Shepherd}, {Pogosyan},
  {Prunet}, {Altamirano}, {Bustos}, {Bronfman}, {Casassus}, {Holzapfel}, {May},
  {Pen}, {Torres}, \& {Udomprasert}}]{Readhead2004ApJ}
{Readhead}, A.~C.~S., {Mason}, B.~S., {Contaldi}, C.~R., {et~al.} 2004, \apj,
  609, 498

\bibitem[{{Refregier}(1999)}]{Refregier1999ASPC}
{Refregier}, A. 1999, in ASP Conf. Ser. 181: Microwave Foregrounds, ed. A.~{de
  Oliveira-Costa} \& M.~{Tegmark}, 219--+

\bibitem[{{Roy}(2006)}]{Roy2006evn}
{Roy}, A. 2006, in Proceedings of the 8th European VLBI Network Symposium.
  September 26-29, 2006,Torun, Poland. Editorial Board: Baan Willem, Bachiller
  Rafael, Booth Roy, Charlot Patrick, Diamond Phil, Garrett Mike, Hong Xiaoyu,
  Jonas Justin, Kus Andrzej, Mantovani Franco, Marecki Andrzej (chairman),
  Olofsson Hans, Schlueter Wolfgang, Tornikoski Merja, Wang Na, Zensus Anton.,
  p.58

\bibitem[{{Scheuer}(1957)}]{Scheuer1957PCPS}
{Scheuer}, P.~A.~G. 1957, in Proceedings of the Cambridge Philisophical
  Society, 764--773

\bibitem[{{Smoot}(1999)}]{Smoot1999ASPC}
{Smoot}, G.~F. 1999, in ASP Conf. Ser. 181: Microwave Foregrounds, ed. A.~{de
  Oliveira-Costa} \& M.~{Tegmark}, 61--+

\bibitem[{{Sunyaev} \& {Zeldovich}(1970)}]{Sunyaev1970ApnSS}
{Sunyaev}, R.~A. \& {Zeldovich}, Y.~B. 1970, \apss, 7, 3

\bibitem[{{Tegmark} {et~al.}(2000){Tegmark}, {Eisenstein}, {Hu}, \& {de
  Oliveira-Costa}}]{Tegmark2000ApJ}
{Tegmark}, M., {Eisenstein}, D.~J., {Hu}, W., \& {de Oliveira-Costa}, A. 2000,
  \apj, 530, 133

\bibitem[{{Weiler} \& {de Pater}(1983)}]{Weiler1983ApJS}
{Weiler}, K.~W. \& {de Pater}, I. 1983, \apjs, 52, 293

\end{thebibliography}

\begin{appendix}

\section{Expected versus observed uncertainties}
\label{sec:ExpectedVsObs}
Here we investigate how the fitted values of $\sigma_{0}$ and $m$
compare with those semi-empirically known. To begin with, the constant
term $\sigma_{0}$ in Eq.~\ref{eq:error} can be decomposed in three
constituents as follows:
\begin{equation}
  \label{eq:sigma0}
  \sigma_{0}^{2}=\sigma_{\mathrm{rms}}^{2} + \sigma_{\mathrm{conf}}^{2}+\sigma_{\mathrm{atm}}^{2}  
\end{equation}
\[
\begin{array}{lp{0.8\linewidth}}
  \sigma_\mathrm{rms}     & the thermal noise, computable from the radiometer formula \\
  \sigma_\mathrm{conf}  & the confusion error, known semi-empirically \citep{Condon1989AJ}   \\
  \sigma_\mathrm{atm}  & variable atmospheric emission error, computable from the atmospheric opacity change.    
\end{array}
\]
\noindent
Separately for each frequency these quantities give:
\begin{table}[h]
  \centering                    
  \begin{tabular}{cccccc} 
    \hline\hline                 
    Frequency  &$\sigma_\mathrm{rms}$ &$\sigma_\mathrm{conf} $  &$\sigma_\mathrm{atm} $ &$\sigma_{0}$ &Fitted $\sigma_{0}$\\    
    (GHz)         &(mJy)                          &(mJy)                                  &(mJy)                                        &(mJy)       &(mJy)\\    
    \\    
    \hline                         
    4.85          &0.16  &0.80  &0.92 &1.23 &1.2  \\
    10.45         &0.22  &0.08  &1.30 &1.32 &1.3  \\
    \hline                                  
  \end{tabular}
\end{table}

\noindent
being satisfactorily close to the expected values. Similarly, the flux
dependent part of Eq.~\ref{eq:error} can be understood as a
multi-factor effect. Specifically, it can be written that:
\begin{equation}
  \label{eq:m}
  m^{2}=m_{\mathrm{poi}}^{2} + m_{\mathrm{cal}}^{2}+m_{\mathrm{atm}}^{2}  
\end{equation}
\[
\begin{array}{lp{0.8\linewidth}}
  m_\mathrm{poi} & pointing offset error, easily calculable on
  the basis of a Gaussian-like beam pattern and from the
  measured average pointing offsets ($\approx 4\arcsec$)\\
  m_\mathrm{cal} & instability of noise diode estimated from Intra-day
  Variability experiments (Kraus priv. comm.)  \\
  m_\mathrm{atm} & variable atmospheric absorption error,
  estimated from the Water Vapor Radiometer data
  \citep{Roy2006evn}.
\end{array}
\]
The expected factor $m'$ as compared to the fitted one $m$, will be:
\begin{table}[h]
  \centering                    
  \begin{tabular}{cccccc} 
    \hline\hline                 
    Frequency  &$m_\mathrm{poi}$ &$m_\mathrm{cal} $  &$m_\mathrm{atm} $ &$m$ &Fitted $m$\\    
    (GHz)         &(\%)                        &(\%)                          &(\%)                         &(\%)       &(\%)\\    
    \\    
    \hline                         
    4.85          &0.21  &1.3  &0.004 &1.32 &1.3  \\
    10.45         &1.01  &1.3  &0.005 &1.64 &1.6  \\
    \hline                                  
  \end{tabular}
\end{table}
\noindent

The term determining the detection threshold is clearly the term
$\sigma_{0}$ in Eq.~\ref{eq:sigma0}. From the above discussion, it is
clear that for the low frequency observations both the atmospheric and
the confusion terms are significant. However, For the higher frequency
there appears a decrease in the confusion term (due to the smaller
beam size) and the dominant remaining factor is the atmospheric
itself.

\section{Resolving the confusion}
\label{app:confusion}
Here we present a method to partially resolve the confusion caused by
known field sources. The goal is to reconstruct the flux density of a
target source whenever possible from knowing the parameters of the
ones causing the confusion.  These parameters are known from the
observations described here since they also belong to the same source
sample and hence are observed. Note that in the majority of the cases
($\sim 73\%$, table~\ref{tab:confflavors}) the de-confusion is not
possible. The reason for that may be either that the confusing sources
are not detected or they are confused themselves.
\begin{figure}
  \centering
  \includegraphics[width=0.4\textwidth]{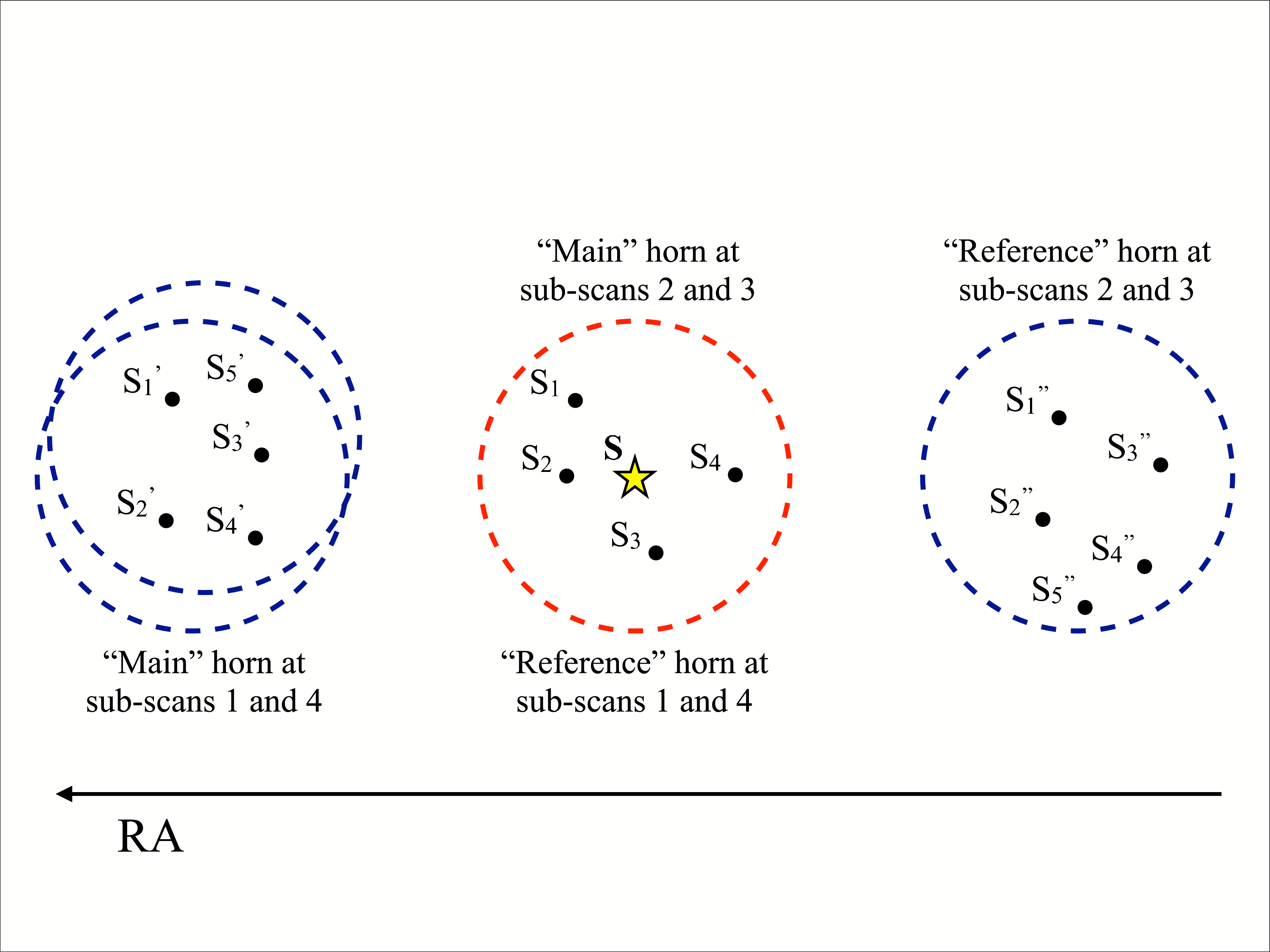}
  \caption{The horn arrangement as a function of time during the
    execution of an observation. In blue is the horn that is pointing
    off-source each time and in red the one on-source. Within each
    horn there might be a population of confusing sources the flux of
    which is represented by $S_{\mathrm{i}}$, ${S_{\mathrm{i}}}'$ or
    ${S_{\mathrm{i}}}''$, with i=1, 2, 3, 4 ... $S$ in yellow is the
    target source. The two blue circles on the left are misplaced
    because the sky rotates within a scan.}
  \label{fig:Confusion2}
\end{figure}
The following analysis is entirely done in the (RA,DEC) space. 

Assume a certain orientation, for instance of the 4.85\,GHz dual beam
system, with respect to the target source and a distribution of
confusing sources as shown in Fig.~\ref{fig:Confusion2}.  There, the
target is represented by the yellow star symbol. In that illustration
there are three distinguishable populations of sources: The ``on''
population, made of sources that lie within a radius of 1 beam-width
($FWHM$) about the target source ($S_{i}$ in
Fig.~\ref{fig:Confusion2}) and each contributes antenna temperature
$T_{i}$. Hence, cumulatively this group will contribute a brightness
temperature:
\begin{equation}
\label{eq:ConfSRCsON}
T_{\mathrm{ON}}=\sum_{\mathrm{i}} T_{\mathrm{i}}
\end{equation}
The ``SUB-1,4'' population, which are the sources $S_{i}^{'}$ located
within a circle of 1 $FWHM$ of the horn position during sub-scan 1 or
4. This is the position of the main horn during the 1$^\mathrm{st}$
and 4$^\mathrm{th}$ sub-scan. This population will contribute a
brightness temperature:
\begin{equation}
\label{eq:ConfSRCs14}
T_{1,4}=\sum_{\mathrm{i}} T_{\mathrm{i}}^{'}
\end{equation}
The ``SUB-2 and 3'' population of the sources occupying the beam
position during sub-scans 2 and 3. This is the position of the
reference horn during those sub-scans. Their contribution will then
be:
\begin{equation}
\label{eq:ConfSRCs23}
T_{2-3}=\sum_{\mathrm{i}} T_{\mathrm{i}}^{''}
\end{equation}
In Eq.~\ref{eq:ConfSRCsON}, \ref{eq:ConfSRCs14} and
\ref{eq:ConfSRCs23} $T_{\mathrm{i}}$, $T_{\mathrm{i}}^{'}$ or
$T_{\mathrm{i}}^{''}$, is the brightness temperature contribution of a
source at the frequency of interest after accounting for the distance
from the center of the beam. Hence, a source of intrinsic brightness
temperature $T_{\mathrm{src}}$ that lies x$_{0}^{''}$ from the center
of the beam of the 4.85\,GHz system, will contribute:
\begin{equation}
\label{eq:TContrib}
T=T_{\mathrm{src}}\cdot e^{-4\,\ln(2)\,\frac{x_{0}^{2}}{FWHM^{2}}}
\end{equation}
From the above it is clear that $T_\mathrm{ON}$, $T_{1-4}$ and
$T_{2-3}$ will be added to the system temperature $T_{\mathrm{sys}}$
altering the result of the differentiation method.

Preserving the notation used in Sect.~\ref{subsec:DataRed}, it can be
shown that the ``real'' source brightness temperature,
$T_{\mathrm{real}}$ can be recovered from observable
$T_{\mathrm{left}}$ and $T_{\mathrm{right}}$, by:
\begin{equation} \label{eq:TLRev}
T_{\mathrm{left}}=T_{\mathrm{real}}+T_{\mathrm{ON}}-\frac{T_{2-3}}{2}-\frac{T_{1}}{2}
\end{equation}
\begin{equation} \label{eq:TRRev}
T_{\mathrm{right}}=T_{\mathrm{real}}+T_{\mathrm{ON}}-\frac{T_{2-3}}{2}-\frac{T_{4}}{2}
\end{equation}
Because the terms $T_{\mathrm{real}}$ and $T_\mathrm{ON}$ in practice
cannot be resolved ({\sl cluster} cases, angular resolution
limitation), it is meaningless to refer to them separately. This is
why we refer to the {\sl cluster} cases separately throughout the text
and why we do not include them in table~\ref{tab:final}. For the sake
of the following discussion we refer to them cumulatively as
$T_{\mathrm{obs}}=T_{\mathrm{real}}+T_{\mathrm{ON}}$.

This simple method has some weaknesses:
\begin{enumerate}
\item{\bf Clustered ``confusers'':} In the above discussion it is
  presumed that the flux densities of the members of a population
  (e.g. $S_{i}^{'}$), are known from the measurement of which target
  was themselves (all the sources we discuss are from of the same
  sample after all and hence have been targeted). This is true only if
  the distance of a pair of sources of the same population is $\ge
  FWHM/2$. Hence, the above method has been applied only in those
  cases.
\item{\bf Missing ``confusers'':} The confusing sources are searched
  among the the NVSS ones. Hence, sources that are not detected by the
  NVSS survey which may become detectable at higher frequencies are
  neglected.
\item{\bf Upper limits:} As seen in table~\ref{tab:final}, often the
  upper limits in the flux density are significant. However, they are
  not accounted for during the de-confusion algorithm.
\item{\bf No corrections applied:} The correction discussed in
  Sect.~\ref{subsec:corrections} are not applied for the confusing
  sources during the resolving algorithm. 
\item{\bf Inaccurate positions of beams and non-Gaussian beams:} In
  all the above it has been assumed that the positions of the beams
  are precisely known and that there are no pointing
  offsets. Furthermore, the beam pattern is supposed to be described
  by a simple circular Gaussian.
\end{enumerate}

\section{Data reduction ``pipeline''}
\label{sec:cbionon}
\begin{figure}
  \centering
  \includegraphics[width=0.4\textwidth]{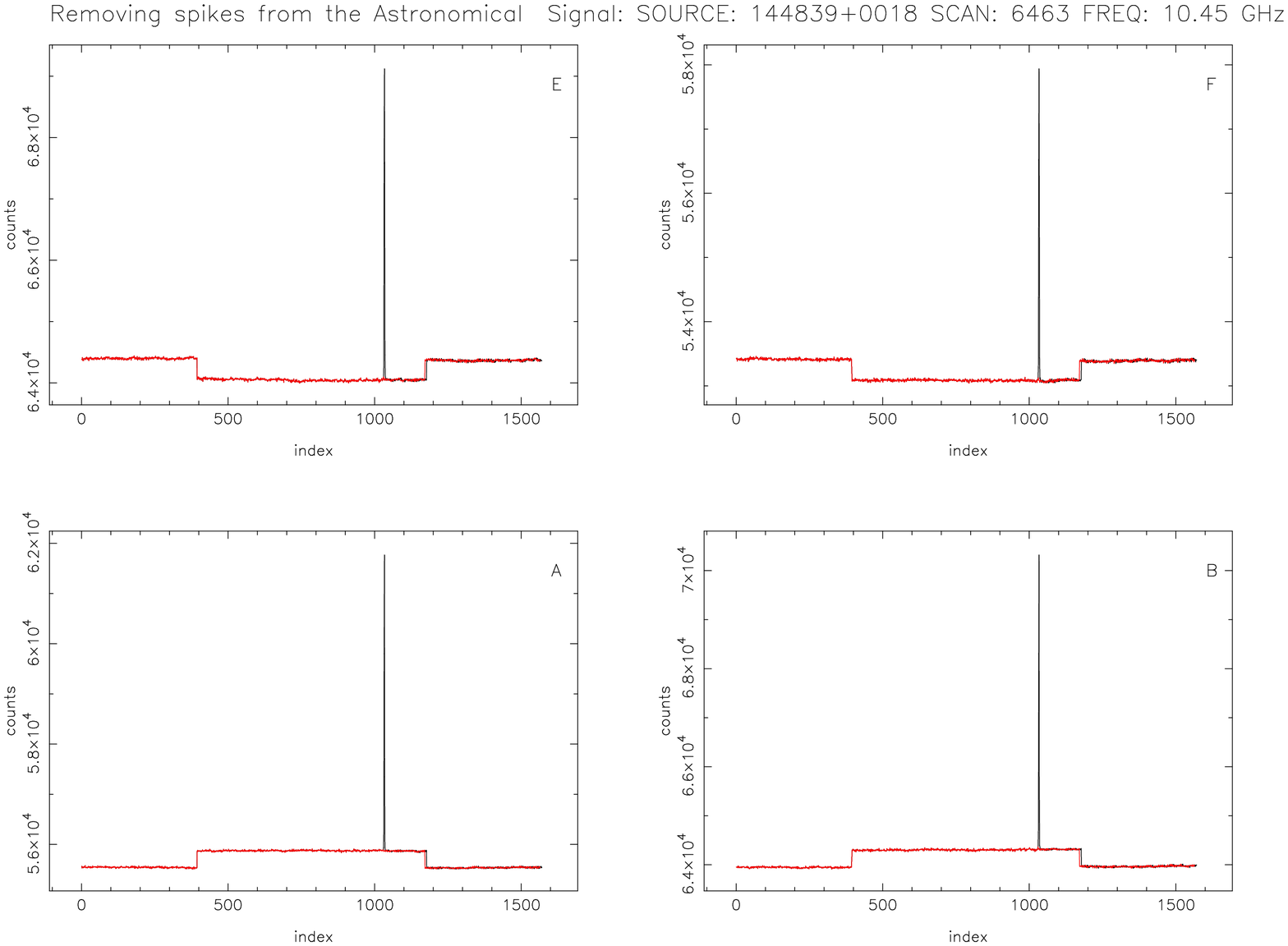}
  \includegraphics[width=0.4\textwidth]{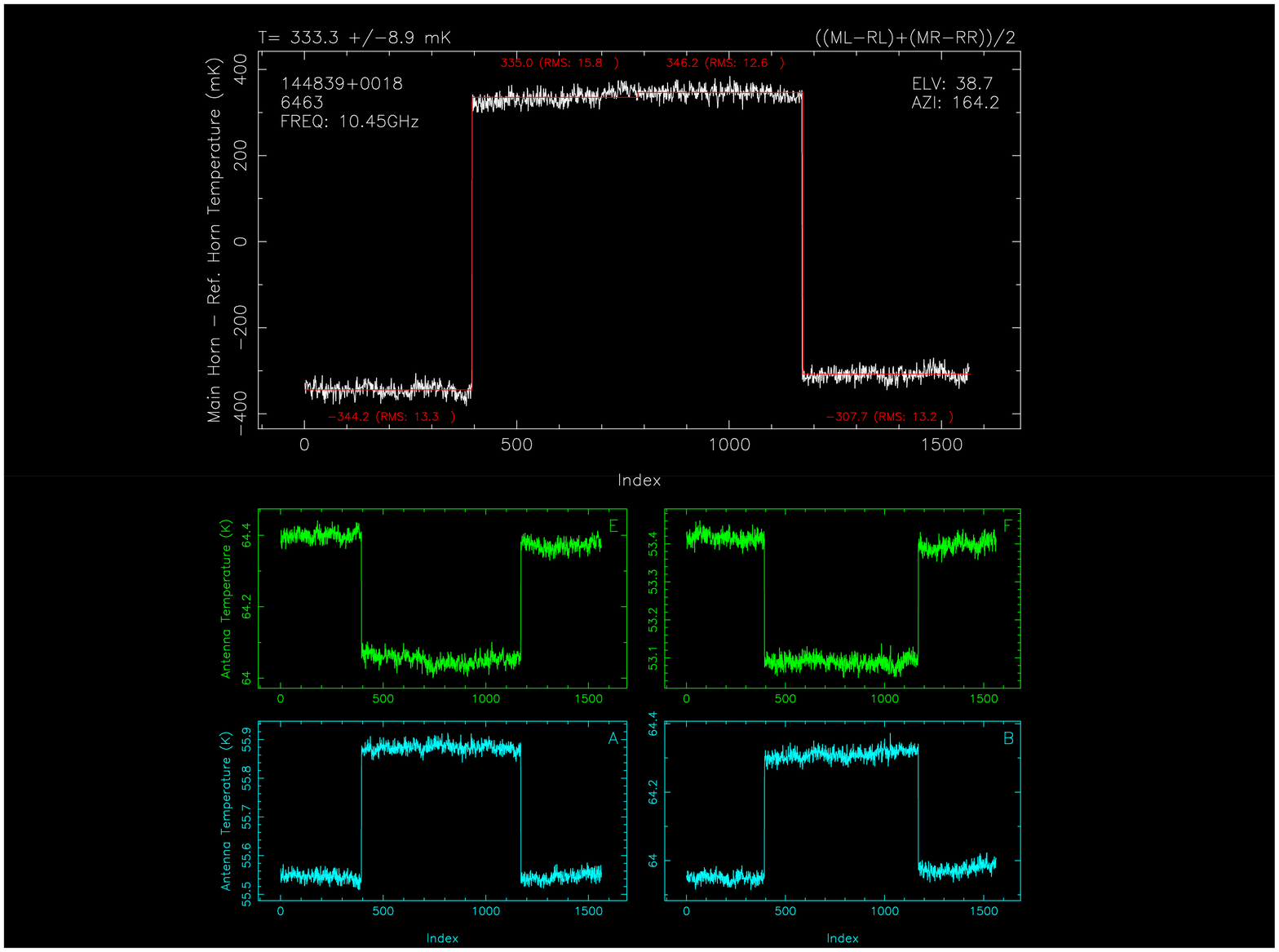}
  \caption{Demonstration of the efficiency of the RFI mitigation
  algorithm. Top panel: the sky signal before (black) and after (red)
  RFI mitigation. 
  Lower panel: the final detection profile free of RFI.}
  \label{fig:RFI} 
\end{figure}

\begin{figure}
  \centering
  \includegraphics[width=0.14\textwidth]{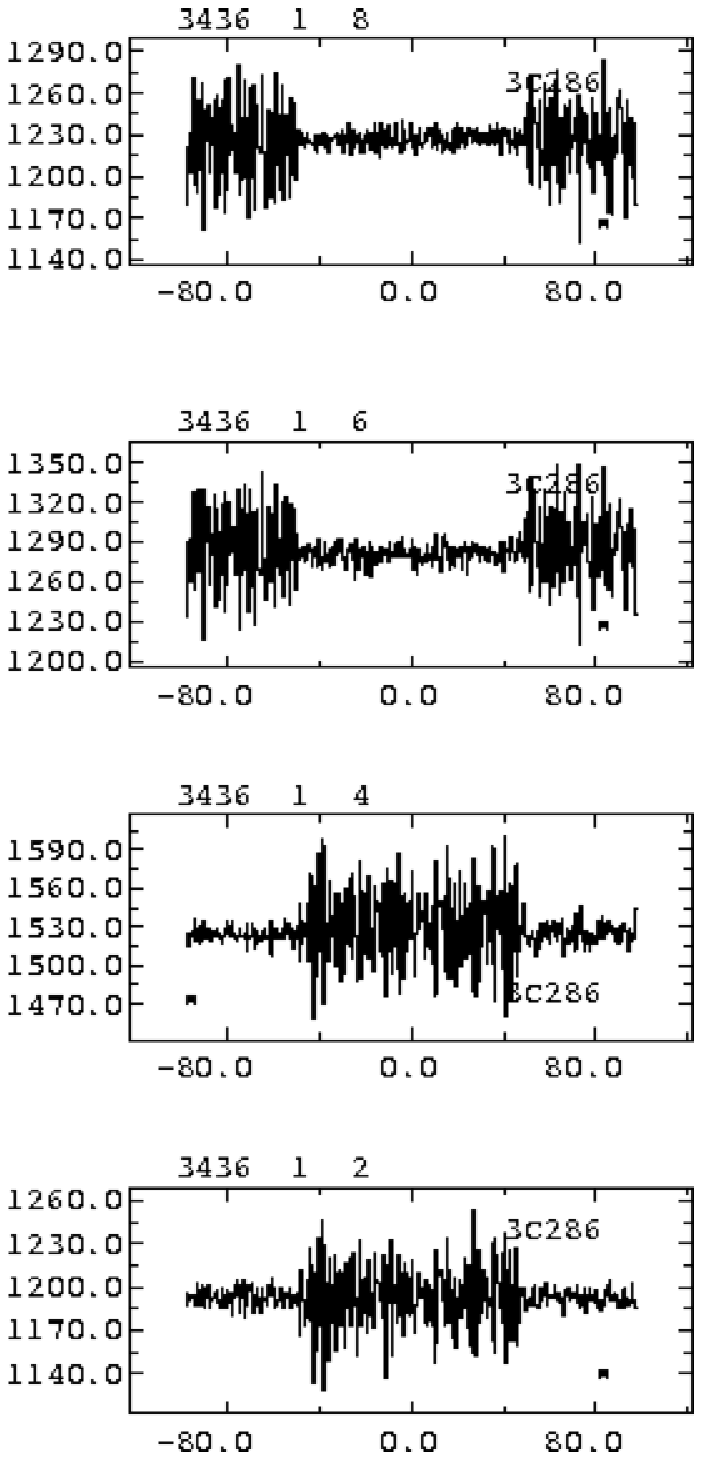}
  \includegraphics[width=0.15\textwidth]{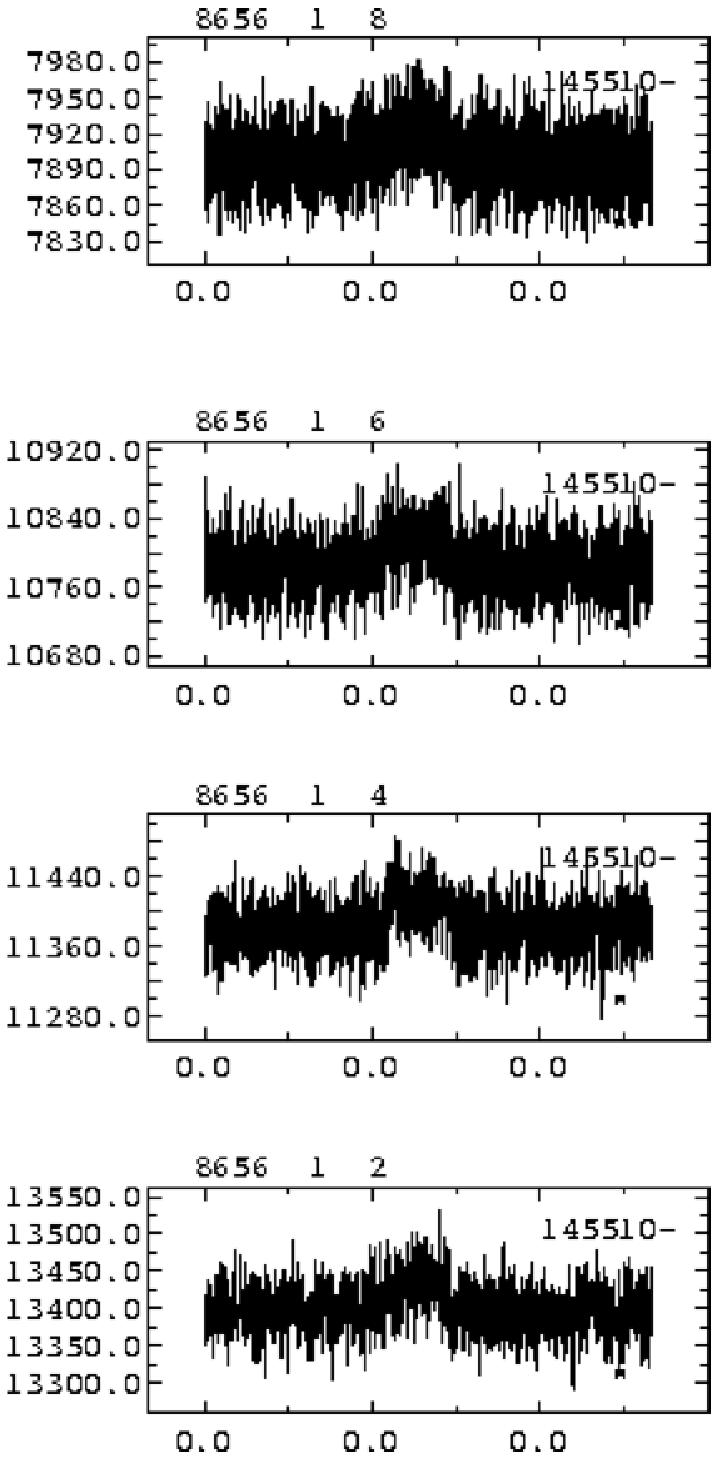}
  \includegraphics[width=0.15\textwidth]{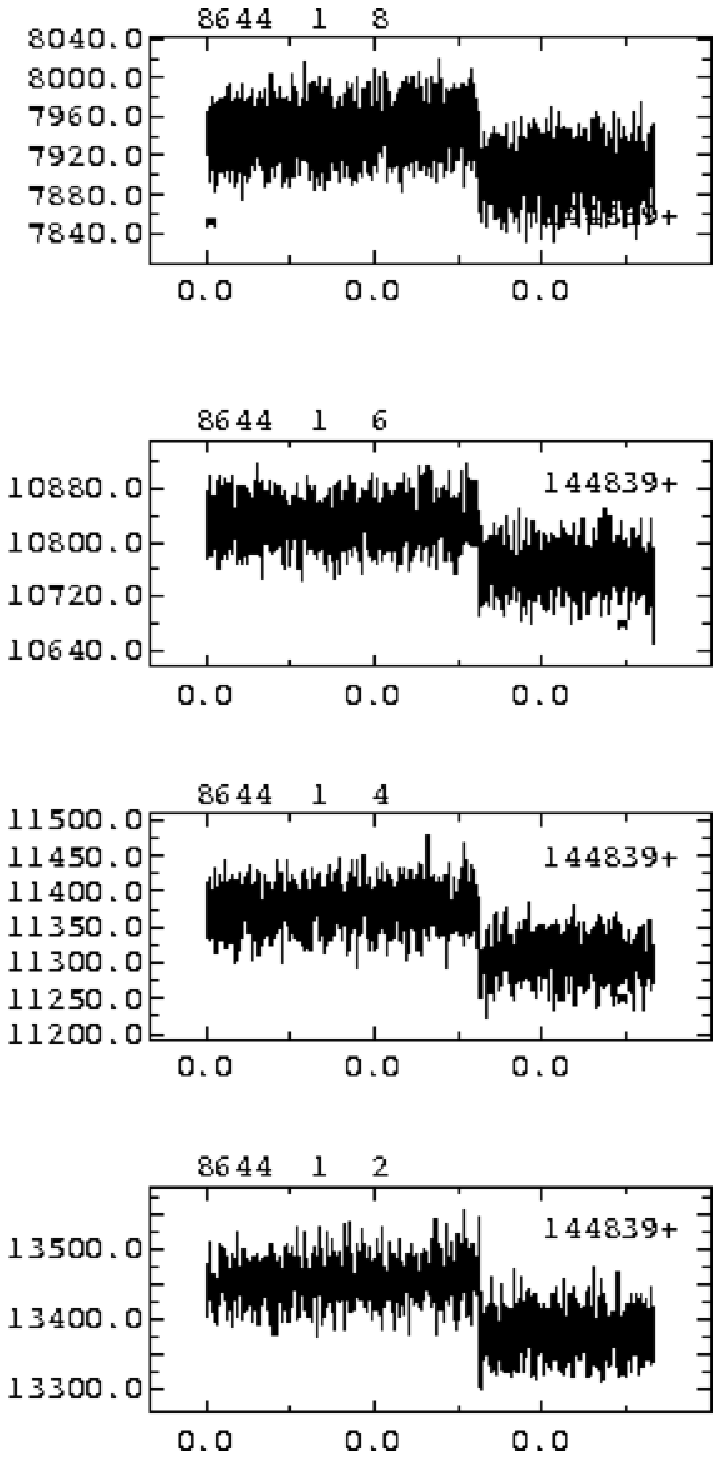}
  \caption{Characteristic cases of intra-scan instabilities of the
  noise diode signal. Each column corresponds to a different scan and
  each row to a different channel. The signal is in terms of counts.}
  \label{fig:Tcalprob} 
\end{figure}

The data volume acquired during the course of the current project has
been reduced in a pipeline manner. Effort has been put into developing
software beyond the standard data reduction packages used in
Effelsberg that could assist the observer to reduce the data as
automatically as possible at all stages. Here we attempt a rough and
very brief description of only some of the steps followed. Throughout
the pipeline, every system parameter is monitored and recorded. Some
details are omitted in this description.
\paragraph{\bf The front-end:}
The front-end of the pipeline is the point at which ``counts'' (power)
from the telescope are piped in the data reduction code. The input
consists of four power data channels two (LCP and RCP) for each horn
along with the signal from a noise diode of known temperature, for
each one of them, i.e. eight channels in total.
\paragraph{\bf RFI mitigation:}
Before any operation is applied to the signal, Radio Frequency
Interference (RFI) mitigation takes place. In Fig.~\ref{fig:RFI} an
example is shown. Here, black represents the signal before and red the
signal after RFI mitigation. In the top panel all four channels of the
sky signal are shown in terms of ``counts''. A short-lived spike of
extremely intense radiation, characteristic of RFI, is clearly seen. A
routine iteratively measures the RMS in that sub-scan and removes the
points above a pre-set threshold. The resulting signal is shown in
red. The same procedure is followed for the noise diode
signal. Finally, the bottom panel shows the final detection pattern
free of RFI.
\paragraph{\bf The signal pre-calibration:}
After the signals have been ``cleaned'' comes the stage of the
comparison of each data point with the noise diode signal, both being
measured in counts. The demand for achieving flux densities as low as
theoretically predicted for the 100\,m telescope imposes the necessity
of having a noise diode signal that ideally should be constant with an
RMS of no more than a fraction of a percentile. However, often
occurring cross-talk between different channels or other effects, may
result in intra-scan instabilities (as shown in
Fig.~\ref{fig:Tcalprob}) that may distort the detection pattern. The
way around this problem has been the idea to normalize ("calibrate")
the data to the average, over the whole scan, diode signal. The
default would be a point-by-point calibration that may on the other
hand significantly distort the detection pattern.
\paragraph{\bf System temperature measurement:} 
Having the data pre-calibrated (meaning in terms of antenna
temperature), allows system temperature measurements. That in turn,
allows measuring the atmospheric opacity for each particular observing
session by using the system temperature of each scan. Later in the
pipeline, this information is used for correcting for the opacity.
\paragraph{\bf The corrections:}
Following the previous stage is that of subtracting the signal of two
feeds and the calculation of the antenna temperature.  Afterwards,the
opacity, gain curve and sensitivity corrections are applied as
described in Sect.~\ref{subsec:corrections}.
\paragraph{\bf The quality check:}
The conceptual end of the pipeline is the quality check subject to
which has been every single scan. The term ``quality check'' wraps up
a number of tests imposed on each scan. Some of them are:
\begin{enumerate}
\item The system temperature of each channel is compared to the
  empirically expected one. Flags are raised at excess of 10, 20 and
  30\,\%. This test serves as an excellent tracer of weather effects,
  system defects etc.
\item A second test is the RMS and the peak-to-peak variation of the
  data in each sub-scan for each channel separately as well as for the
  final profile. An increase can be caused by extreme non-linear
  atmospheric effects as well as linear slopes present in the
  data. The latter is most often the result of increasing atmospheric
  opacity as the source is tracked at low elevations.
\item In order to trace cases that show a clear linear drift as a
  result of increasing opacity, each scan has been sliced into four
  segments. A straight line has consecutively been fitted to each
  segment. A flag is raised when the slope of a segment is above some
  preset value.
\item It is examined whether the final measurement profile is inverted
  and if so whether the absolute source flux density satisfies the
  detection threshold being set. It is possible in cases of confusion
  that a source in the off position may result in an inverted profile.
\item It is checked whether sensitivity factor (K/Jy) applied to a
  scan agrees with the empirically expected value with a tolerance of
  5\,\%.
\end{enumerate}

\end{appendix}

\end{document}